\documentstyle[12pt,epsfig]{article}
\def\np#1#2#3  {{\it Nucl. Phys. }{\bf #1} (19#3) #2} 
\def\nc#1#2#3  {{\it Nuovo. Cim. }{\bf #1} (19#3) #2} 
\def\pl#1#2#3  {{\it Phys. Lett. }{\bf #1} (19#3) #2} 
\def\pr#1#2#3  {{\it Phys. Rev. }{\bf #1} (19#3) #2} 
\def\prl#1#2#3  {{\it Phys. Rev. Lett. }{\bf #1} (19#3) #2} 
\def\prep#1#2#3 {{\it Phys. Rep. }{\bf #1} (19#3) #2} 
\def\zp#1#2#3  {{\it Z. Phys. }{\bf #1} (19#3) #2} 
\def\rmp#1#2#3  {{\it Rev. Mod. Phys. }{\bf #1} (19#3) #2} 
\def\epj#1#2#3  {{\it Eur. Phys. J. }{\bf #1} (19#3) #2} 
\def\hepph  #1 {{\tt hep-ph/#1}}

\begin{document}

\def\tr{\mathop{\rm tr}}
\def\Tr{\mathop{\rm Tr}}
\def\Im{\mathop{\rm Im}}
\def\Re{\mathop{\rm Re}}
\def\bR{\mathop{\bf R}{}}
\def\bC{\mathop{\bf C}{}}
\def\C{\mathop{\rm C}}
\def\bra#1{\left\langle #1\right|}
\def\ket#1{\left| #1\right\rangle}
\def\VEV#1{\left\langle #1\right\rangle}
\def\gdot#1{\rlap{$#1$}/}
\def\abs#1{\left| #1\right|}
  \newcommand{\ccaption}[2]{
    \begin{center}
    \parbox{0.85\textwidth}{
      \caption[#1]{\small{\it{#2}}}
      }
    \end{center}
    }
\def\beq{\begin{equation}}
\def\eeq{\end{equation}}
\def\eq{\beq\eeq}
\def\beqn{\begin{eqnarray}}
\def\eeqn{\end{eqnarray}}
\relax
\let\h=\hat
\newcommand\sss{\scriptscriptstyle}
\newcommand\gs{g_{\sss S}}
\newcommand\as{\alpha_{\sss S}}         
\newcommand\ep{\epsilon}
\newcommand\Th{\theta}
\newcommand\epb{\overline{\epsilon}}
\newcommand\aem{\alpha_{\rm em}}
\newcommand\refq[1]{$^{[#1]}$}
\newcommand\avr[1]{\left\langle #1 \right\rangle}
\newcommand\lambdamsb{\Lambda_5^{\rm \sss \overline{MS}}}
\newcommand\qqb{{q\bar{q}}}
\newcommand\qb{\bar{q}}
\newcommand\xto{\tilde{x}_1}
\newcommand\xtt{\tilde{x}_2}
\newcommand\aoat{a_1 a_2}
\newcommand\Oop{{\cal O}}
\newcommand\Sfun{{\cal S}}
\newcommand\Pfun{{\cal P}}
\newcommand\mug{\mu_\gamma}
\newcommand\mue{\mu_e}
\newcommand\muf{\mu_{\sss F}}
\newcommand\mufp{\mu_{\sss F}^\prime}
\newcommand\mufs{\mu_{\sss F}^{\prime\prime}}
\newcommand\mur{\mu_{\sss R}}
\newcommand\murp{\mu_{\sss R}^\prime}
\newcommand\murs{\mu_{\sss R}^{\prime\prime}}
\newcommand\muh{\mu_{\sss H}}
\newcommand\muhp{\mu_{\sss H}^\prime}
\newcommand\muhs{\mu_{\sss H}^{\prime\prime}}
\newcommand\muo{\mu_0}
\newcommand\MSB{{\rm \overline{MS}}}
\newcommand\DIG{{\rm DIS}_\gamma}
\newcommand\CA{C_{\sss A}}
\newcommand\DA{D_{\sss A}}
\newcommand\CF{C_{\sss F}}
\newcommand\TF{T_{\sss F}}
\newcommand\pt{p_{\sss T}}
\newcommand\kt{k_{\sss T}}
\newcommand\ptg{p_{{\sss T}\gamma}}
\newcommand\xtg{x_{\sss T}^\gamma}
\newcommand\etag{\eta_\gamma}
\newcommand\phig{\phi_\gamma}
\newcommand\ptj{p_{{\sss T}j}}
\newcommand\etaj{\eta_j}
\newcommand\epg{\epsilon_\gamma}
\newcommand\epc{\epsilon_c}
\newcommand\epem{e^+e^-}
\def\rightrightarrows{\rlap{\lower 2.5 pt \hbox{$\mathchar\rightarrow$}} 
                      \raise 1pt \hbox {$\mathchar\rightarrow$}}
\def\rightleftarrows{\rlap{\lower 2.5 pt \hbox{$\mathchar\leftarrow$}} 
                     \raise 1pt \hbox {$\mathchar\rightarrow$}}
\begin{titlepage}
\nopagebreak
\flushright{
        \begin{minipage}{4cm}
        CERN-TH/99-247\\
        YITP-99-47 \hfill \\
        hep-ph/9908387\hfill \\
        \end{minipage}        
}
\vfill
\begin{center}
{\large\sc Isolated-photon production\\ 
in polarized $pp$ collisions}
\vskip .5cm
{\bf Stefano Frixione}
\\                    
\vskip .1cm
{CERN, TH Division, Geneva, Switzerland}\\
\vskip .5cm
{\bf Werner Vogelsang}
\\                    
\vskip .1cm
{C.N.\ Yang Institute for Theoretical Physics\\
State University of New York at Stony Brook\\
Stony Brook, NY 11794-3840, USA}\\
\end{center}
\nopagebreak
\vfill
\begin{abstract}
\noindent
We perform a detailed study of the production of isolated prompt
photons in polarized hadronic collisions, in the centre-of-mass
energy range relevant to RHIC. We compare the results obtained 
for a traditional cone-isolation prescription, with those obtained 
by imposing an isolation condition that eliminates any contribution to the
cross section from the fragmentation mechanism. The latter prescription will 
allow us to present the first fully consistent next-to-leading order 
calculation in polarized prompt-photon production. We will discuss the 
theoretical uncertainties affecting the cross section, addressing
the issue of the reliability of the perturbative expansion, for
both inclusive isolated-photon and photon-plus-jet observables. Finally,
we will study the dependence of our predictions upon the 
polarized parton densities, and the implications for the 
measurability of the gluon density.

\end{abstract}        
\vfill
CERN-TH/99-247\hfill~\\
August 1999\hfill ~
\end{titlepage}

\section{Introduction\label{sec:intro}}

A direct measurement of the gluon distribution in the proton is both
very interesting and very difficult. This is true, in particular,
for the spin-dependent gluon density $\Delta g$ of a longitudinally 
polarized proton. So far, the only information on the nucleon
polarized parton distributions comes from polarized deep-inelastic 
lepton--nucleon scattering (DIS). Here, in principle, $\Delta g$ 
could be determined from scaling violations; however, in practice 
this procedure is afflicted by very large uncertainties due to the limited
accuracy of the data and, in particular, to the fact that so far only 
{\em fixed-target} DIS experiments have been carried out in the polarized 
case, which, consequently, have a rather limited lever arm in $Q^2$. 
Thus, the proton spin-dependent gluon density $\Delta g$ is experimentally 
constrained only very little so far. 

In order to perform a direct determination of the gluon
distribution, one needs to consider physical processes, which are 
predominantly initiated by gluons at the parton level; the contamination
from quark-initiated subprocesses must be under good theoretical
control; and finally, the process must take place at a measurable
rate, taking experimental efficiencies into proper account. Only a few
processes are known that meet all these requirements. One example is
the production of heavy flavours in photon-hadron collisions; the
cross section for this process in the case of polarized scattering
has recently been computed~\cite{HVQpol} in QCD at next-to-leading 
order (NLO). The COMPASS collaboration at CERN~\cite{COMPASS}, and possibly 
even HERA in the polarized configuration~\cite{heracharm}, will be able 
to exploit charm production in order to constrain the polarized gluon
density. Such measurements are on the other hand severely limited 
by the low experimental efficiency of charmed-meson tagging.

A second possibility is given by jet hadro- and photoproduction; also in 
this case, QCD cross sections for polarized collisions are now known to 
NLO~\cite{dfsv,dff}. It will indeed be attempted to determine $\Delta g$, 
at the forthcoming polarized RHIC $pp$ collider~\cite{rhic}, by a 
measurement of the spin asymmetry in jet production, and the recent 
study of ref.~\cite{dfsv} has demonstrated that this approach has 
very promising prospects.

In the unpolarized case, the classical tool for determining the 
gluon density at intermediate and large $x$ has been prompt-photon 
production, $pp\rightarrow \gamma X$ and $pN \rightarrow \gamma X$, 
in fixed-target experiments~\cite{photondata}. Indeed, data on prompt 
photons have been the backbone of the gluon determination in many analyses 
of parton densities. The main reason for this is that, at leading order, 
a photon in the final state is produced in the reactions 
$qg\to\gamma q$ and $q\bar{q}\to\gamma g$, with the contribution of the 
former subprocess being obviously sensitive to the gluon and usually 
dominant over that of the latter. It is the `point-like' coupling of 
the photon to the quark in these subprocesses that is responsible
for a much cleaner signal than, say, for the inclusive production of
a $\pi^0$, which proceeds necessarily through a fragmentation process.

The cleanliness of the signal is of course an advantage that also 
counts at colliders. The aim of this paper is to provide a detailed
study, performed in perturbative QCD at NLO accuracy, of the 
production mechanism for prompt photons at {\it polarized} hadronic
colliders, such as RHIC. In the rest of this introduction, we will
briefly review the present theoretical knowledge of prompt-photon 
production. In particular, we will explain why, in our opinion,
a theoretical reappraisal of this matter is needed, before the 
data-taking will start at RHIC.

Prompt-photon data obtained at hadronic colliders have been 
used as a constraint for the unpolarized gluon density~\cite{photondata}.
Thinking of the polarized case again, it is clear that the production of 
photons with polarized beams at RHIC is likely to be a very promising source 
of information on $\Delta g$. We note that having $pp$ reactions 
(as opposed to $p\bar{p}$ ones as hitherto in the unpolarized case) 
is also an advantage, since the competing LO subprocess $q\bar{q}\to\gamma g$ 
does not receive any contributions from valence--valence scattering here. 
Compared to jets, the prompt-photon reaction shows a larger spin asymmetry, 
even though of course the jet rate is much higher at a given $p_T$, resulting 
in smaller statistical errors. Eventually, both reactions, plus also charm 
production and $\pi^0$ production in polarized $pp$ scattering, 
can be utilized to constrain $\Delta g$ at RHIC, and it 
will be interesting to see how far the various channels will provide 
compatible pieces of information, and/or whether they will complement one 
another.

Unfortunately, the cleanliness of the prompt-photon signal alleged above
is limited. As is well known, photons can also be produced through
a fragmentation process, in which a parton, scattered or produced in
a QCD reaction, fragments into a photon plus a number of hadrons.
The problem with the fragmentation 
component in the prompt-photon reaction is twofold: first, it 
brings us back to the dependence on non-perturbative fragmentation
functions, similar to the case of pion production addressed above,
even though for prompt-photon production of course only a certain part 
of the total signal depends on the fragmentation functions. So far, the 
photon fragmentation functions are only insufficiently known; first 
information is emerging from the LEP experiments~\cite{lepphoton}. Secondly, 
{\em all} QCD partonic reactions contribute to the fragmentation component;
thus the advantage of having a priori only one partonic reaction 
($q\bar{q}\to\gamma g$) competing with the signal ($qg\to\gamma q$) 
is lost, even though some of the subprocesses relevant to the 
fragmentation part at the same time result from a gluon initial state.

Numerical studies~\cite{acfgp,ggrv,vv,afgkpw} for photon production in 
unpolarized collisions, based on predictions~\cite{acfgp,grvfrag,bfg}
for the photon fragmentation 
functions that turned out to be compatible with the sparse LEP data, 
demonstrate that the fragmentation component is not messing up things too 
much, even though it cannot be neglected in a careful study. In the 
fixed-target 
regime, it amounts to an effect of about $20\%$. At collider energies, 
it would easily make up for about half of the observed photons; however, here
the situation is saved by the so-called `isolation' cut, which is
imposed on the photon signal in experiment. Isolation is an experimental
necessity: in a hadronic environment the study of photons in the final state 
is complicated by the abundance of $\pi^0$'s, eventually decaying into pairs 
of $\gamma$'s. The isolation cut simply serves to improve the signal-to-noise 
ratio: if a given neighbourhood of the photon is free of
energetic hadron tracks, the event is kept; it is rejected otherwise. In 
principle, there is a large freedom in the choice of specific isolation cuts, 
the only requirement being that they must strongly suppress the background 
$\pi^0\to\gamma\gamma$, while keeping the signal at a measurable rate.
Traditionally, isolation is realized by drawing a cone of fixed aperture
in azimuthal angle--pseudorapidity space around the photon, and by 
restricting the hadronic transverse energy allowed in this cone to a certain
fraction (of the order of less than $10\%$) of the photon transverse energy. 
In this way, 
it is clear that the fragmentation contribution, resulting from an 
essentially collinear process, will be diminished~\cite{bq1}. 
In actual numbers, it is not expected~\cite{ggrv,vv} that it will
remain responsible for more than 15--20\% of the photon signal 
after isolation.

Studies of the backgrounds to prompt-photon production expected for
RHIC have been reported in refs.~\cite{goto,bland,gull}, based on
parton-shower Monte-Carlo event generators. It is
anticipated~\cite{goto,bland} that isolation cuts will also have to be
applied in prompt-photon measurements at polarized RHIC. When working
out theory predictions for the (un)polarized cross sections and the
resulting spin asymmetry, it is crucial that the calculation properly
treats all effects mentioned so far, in particular those related to
fragmentation and isolation.

It is the objective of this paper to do just this. Of course, 
it must be pointed out that several studies with a similar 
focus have been presented before. In particular, after the basic idea
of determining $\Delta g$ in $\vec{p}\vec{p}\to \gamma X$ had been formulated
and developed in ref.~\cite{bq}, the QCD corrections
to the `direct' (i.e. {\em non}-fragmentation) component of polarized 
prompt-photon production were first calculated in refs.~\cite{cont,gv} and
applied for phenomenological predictions in refs.~\cite{cont1,gv1}. 
There, the fragmentation component was neglected altogether, and no 
isolation cut was imposed. The calculations presented in refs.~\cite{cont,gv} 
are fully analytical and can be used only for calculating the 
single-inclusive photon cross section, and not for looking at, say, 
photon-plus-jet final states. Still, it was demonstrated~\cite{gv2} 
how the effects of the isolation cut can be implemented into these 
calculations in the approximation that the isolation cone be rather narrow. 
Much more complete phenomenological studies were presented for the case of 
inclusive photons in ref.~\cite{lg} and for $\vec{p}\vec{p} \to \gamma+jet+X$ 
in ref.~\cite{lg1}. 
In these papers, a Monte-Carlo code for the NLO corrections to 
the direct part of the cross section was developed and employed,
which readily allows the isolation constraints to be taken into account.
The results presented in refs.~\cite{lg,lg1} suffer from the fact that the 
fragmentation component to the polarized cross section could be treated 
at the LO level only, since the corresponding NLO corrections 
had not yet been calculated. This
is potentially hazardous, since beyond LO the direct as well as the
fragmentation part of the cross section depends on the factorization
convention chosen in subtracting the final-state collinear singularities.
Only their sum is free of scheme-dependence and is physical, which implies
that a fully consistent NLO calculation affords knowledge of both 
production mechanisms at NLO, even though using a LO fragmentation
contribution instead of the NLO one presumably introduces only a 
minor error from a numerical point of view. 
We emphasize that this situation has not improved since 
then, and it is not the purpose of this paper to do this. Still, this 
paper will present the first {\em fully consistent} NLO calculation of 
polarized prompt-photon production. 

We believe that we have several good reasons for presenting a further
phenomenological study on prompt photons at RHIC. First of all, 
our predictions will be based on the use of a new type of photon isolation 
constraint, introduced recently by one of us~\cite{stefano}, which 
has the virtue of entirely eliminating the unwanted fragmentation 
component to the cross section. 
The basic idea of ref.~\cite{stefano} is not only to restrict 
the total hadronic energy falling into the isolation cone, but to allow 
less and less hadronic energy, the closer to the photon it is deposited,
until eventually no energy at all is allowed exactly collinear to 
the photon. In this way, no collinear configuration is possible, 
and fragmentation does not contribute to the cross section. 
This feature allows us to present complete and theoretically consistent 
NLO predictions for polarized prompt-photon production. This will be done
in terms of a dedicated Monte-Carlo program, which implements the
new type of isolation in both polarized and unpolarized hadronic
collisions. The isolation constraint we promote should be straightforwardly 
implementable in experiment. Here, we have in mind in particular the PHENIX 
detector at RHIC, with the very fine granularity of its electromagnetic 
calorimeter~\cite{goto}; also the STAR detector~\cite{bland}, with its much 
larger angular coverage and its ability to see jets, appears to offer 
promising prospects. 

Another motivation for our study is to look in more detail at the main
theoretical uncertainties in the calculation, resulting from the scale 
dependence of the results. Furthermore, previous theoretical studies have not 
really sufficiently addressed the question of the experimentally achievable 
statistical accuracy in the various conceivable measurements of the spin 
asymmetries in prompt-photon production. This will also be done in
this paper. Here, we somewhat disagree with some previous conclusions 
concerning the usefulness of inclusive measurements compared with
photon-plus-jet ones.

Presenting a study on prompt photons, we cannot ignore a disagreeable
development found in the {\em un}polarized case in the last few years
concerning the comparison between theory predictions and data.
While things worked out very well in the first decade or so of prompt-photon 
experiments, the agreement with the more recent, and the most precise, 
data sets~\cite{cdf} is rather poor and sometimes so bad that the situation
cannot possibly be saved even by `fine'-tuning the unpolarized gluon density!
Clearly, if this situation persists, we will have to worry about whether 
one can really interpret, in the polarized case, future RHIC data 
straightforwardly in terms of $\Delta g$. A possible remedy
for this trend has been brought forward in terms of a smearing of the
transverse momenta of the initial partons participating in the hard 
scattering~\cite{kteff,MRST}, required to be substantially 
larger than what is already introduced by the NLO calculation. 
This approach still remains to be set on a more solid foundation -- 
eventually it should be accounted for to some extent by a $\kt$-resummation 
calculation~\cite{lai} with perturbative as well as non-perturbative 
components. Furthermore, threshold resummations~\cite{thresh}, aiming at the 
high-$\pt$ end, have been shown~\cite{thrphen} to lead to a certain 
improvement in the fixed-target regime. We also note that possible 
inconsistencies between the various data sets 
themselves have been pointed out~\cite{afgkpw}. It remains to be seen
whether or not the agreement between data and theory will be in better shape 
by the time RHIC will perform the first measurements on 
\mbox{$\vec{p}\vec{p}\rightarrow \gamma X$}, 
as a result of the present experimental and, in particular, 
theoretical efforts in this field. Whatever the solution will eventually be, 
NLO theory, as employed in this work, will certainly be an indispensable 
part of it; it should then be straightforward to implement the lessons
from the unpolarized case to the polarized one. It should not be forgotten 
either that RHIC itself should be able to provide new and complementary 
information also in the unpolarized case -- never before have prompt-photon 
data been taken in $pp$ collisions at energies as high as 
\mbox{$\sqrt{S}=$200--500~GeV}.

The remainder of the paper is organized as follows. In
section~\ref{sec:formalism} we will provide the framework for our
calculations. The new isolation definition will be presented in more
detail, as well as the main ingredients for our Monte-Carlo code. The
remaining sections are devoted to numerical studies for RHIC.
Section~\ref{sec:si} deals with the single-inclusive cross section.
In subsection~\ref{subsec:si1} we focus on issues related to
perturbative stability, theoretical uncertainties and the effects
of isolation. Subsection~\ref{subsec:si2} is devoted to studies 
of spin asymmetries for the prompt-photon process at RHIC and
of the sensitivity to $\Delta g$. In section~\ref{sec:corr} we consider 
more differential variables, such as photon-plus-jet ones. Here, 
subsection~\ref{subsec:prel} discusses some general features
of photon-plus-jet observables and also addresses 
their perturbative stability, while subsection~\ref{subsec:pjphen}
presents phenomenological results. Finally, we summarize our work in 
section~\ref{sec:concl}.
\newpage

\section{Isolated photons in perturbative QCD\label{sec:formalism}}
\subsection{Isolation prescriptions}

The production of isolated photons in hadronic collisions can be
written in perturbative QCD as follows
\beqn
&&d\sigma_{AB}(K_A,K_B;K_\gamma)=
\nonumber \\*&&\phantom{aa}
\int dx_1 dx_2 f^{(A)}_a(x_1,\muf) f^{(B)}_b(x_2,\muf) 
d\hat{\sigma}_{ab,\gamma}^{isol}(x_1 K_A,x_2 K_B;K_\gamma;\mur,\muf,\mug)
\nonumber \\*&&
+\int dx_1 dx_2 dz f^{(A)}_a(x_1,\mufp) f^{(B)}_b(x_2,\mufp) 
d\hat{\sigma}_{ab,c}^{isol}(x_1 K_A,x_2 K_B;K_\gamma/z;\murp,\mufp,\mug) 
D^{(c)}_\gamma (z,\mug),\phantom{aaaa}
\label{factth}
\eeqn
where $A$ and $B$ are the incoming hadrons, with momenta $K_A$ and $K_B$
respectively, and a sum over the parton indices $a$, $b$ and $c$ is 
understood. In the first term on the RHS of eq.~(\ref{factth}), 
denoted as the direct component, the subtracted (factorized) 
partonic cross sections 
\mbox{$d\hat{\sigma}_{ab,\gamma}^{isol}$} get contributions from all 
the diagrams with a photon leg. On the other hand, the subtracted
partonic cross sections \mbox{$d\hat{\sigma}_{ab,c}^{isol}$}
appearing in the second term on the RHS of 
eq.~(\ref{factth}) (denoted as the fragmentation component), get
contribution from the pure QCD diagrams, with one of the partons
eventually fragmenting in a photon, in a way described by the 
(perturbatively uncalculable but universal) parton-to-photon
fragmentation function $D^{(c)}_\gamma$. Equation~(\ref{factth}) is
to be regarded as a generic expression for the cross section:
it will apply to unpolarized as well as polarized cross sections;
in the latter case one simply has to substitute the parton densities
$f_i^{(h)}$ and the partonic cross sections 
\mbox{$d\hat{\sigma}_{ab,r}^{isol}$} with their spin-dependent counterparts,
\mbox{$\Delta f_i^{(h)}$} and \mbox{$d\Delta\hat{\sigma}_{ab,r}^{isol}$}
respectively.
Note, however, that the parton-to-photon fragmentation functions
$D^{(c)}_\gamma$ are always the unpolarized ones since we are not 
measuring the polarization of the produced photon. 

As the notation in eq.~(\ref{factth}) indicates, the isolation condition is 
embedded into the partonic cross sections. As mentioned in the introduction,
for all the isolation conditions
known at present, except that of ref.~\cite{stefano}, 
as well as for the case of totally
inclusive (non-isolated) photon production, neither the direct nor
the fragmentation components are {\it separately} well defined
at any fixed order in perturbation theory: only their sum is
physically meaningful. In fact, the direct component is affected
by quark-to-photon collinear divergences, which are
subtracted by the bare fragmentation function that appears in
the unsubtracted fragmentation component. Of course, this subtraction 
is arbitrary as far as finite terms are concerned. This is formally 
expressed in eq.~(\ref{factth}) by the presence of the same scale 
$\mug$ in both the direct and fragmentation components: a finite piece
may be either included in the former or in the latter, without affecting
the physical predictions. The need for introducing a fragmentation 
contribution is physically better motivated from the fact that a QCD hard 
scattering process may produce, again through a fragmentation process, 
a $\rho$ meson that has the same quantum numbers as the photon and can 
thus convert into a photon, leading to the same signal. 

Owing to the presence of the fragmentation remnants, which surround the
emitted photon, the effect of the isolation cuts will be 
a stronger suppression of the fragmentation component relative to
the direct component, with respect to the case of totally inclusive 
photon production. Since the parton-to-photon fragmentation functions
are extremely poorly known, one may adopt two opposite points of view.
\begin{itemize}
\item Define the isolation cuts in order to suppress as much as possible
the fragmentation component. The resulting cross section will be useful
to measure the incoming gluon density or to test the predictions of
the underlying theory. In this context, the unknown fragmentation
functions are regarded as uncertainties affecting the theoretical
predictions.
\item Define the isolation cuts in order to keep a non-negligible
contribution from the fragmentation component. The comparison between
data and the resulting cross section will eventually be used
to extract the parton-to-photon fragmentation functions. This strategy
makes most sense if the initial state is as clean as possible,
which is the case for $\epem$ collisions. 
\end{itemize}

The former criterion leads to the so-called cone approach~\cite{bq1,gv2,cone}.
After tagging the photon, one draws a cone of half-angle $R_0$
around it. The word `cone' can be misleading, being motivated
by $\epem$ physics. Here, the cone is drawn in the pseudorapidity--azimuthal
angle plane, and corresponds to the set of points
\beq
{\cal C}_{R_0}=\left\{(\eta,\phi)\mid
\sqrt{(\eta-\etag)^2+(\phi-\phig)^2}\le R_0\right\},
\label{coneRz}
\eeq
where $\etag$ and $\phig$ are the pseudorapidity and azimuthal
angle of the photon, respectively. The quantity in eq.~(\ref{coneRz})
is boost-invariant, and is therefore suited to be used in collider 
physics. For the photon to be defined as isolated, the total amount
of hadronic transverse energy $E_{T,had}(R_0)$ found in this cone 
must fulfil the following condition:
\beq
E_{T,had}(R_0)\le \epc \ptg,
\label{iscondA}
\eeq
where $\epc$ is a small number, and $\ptg$ is the transverse momentum
of the photon. This isolation prescription was proven to be infrared safe 
at all orders of perturbation theory in ref.~\cite{CFP}. 
The smaller $\epc$, the tighter the isolation. Loosely
speaking, for vanishing $\epc$ the direct component behaves like
\mbox{$\log\epc$}, while the fragmentation component behaves like
\mbox{$\epc\log\epc$}. Thus, for $\epc\to 0$ eq.~(\ref{factth})
diverges. This is obvious since the limit $\epc\to 0$ corresponds
to a fully-isolated cross section, which cannot be a meaningful
quantity, whether experimentally (because of limited energy resolution)
or theoretically (because there is no possibility for soft particles 
to be emitted into the cone).

On the other hand, if one actually 
wants to measure the fragmentation functions,
then the so-called democratic approach should be adopted~\cite{dem}. 
The basic idea here is to treat the photon as a QCD parton
in a jet-clustering algorithm, and then to impose a cut on the hadronic 
energy contained in the `jet', which also contains the photon.
This approach has so far been used only in $\epem$ physics,
and we will not discuss it any further in this paper.

In the spirit of the cone approach, an alternative definition of the isolated 
photon has been proposed~\cite{stefano}. After drawing a cone of half-angle 
$R_0$ around the photon axis, all the cones of half-angle $R\le R_0$ are 
considered; their definition is identical to the one given in 
eq.~(\ref{coneRz}), with $R_0$ replaced by $R$. Denoting by
$E_{T,had}(R)$ the total amount of hadronic transverse energy found 
in each of these cones, the photon is isolated if the following 
inequality is satisfied:
\beq
E_{T,had}(R)\le \epg\ptg {\cal Y}(R),
\label{iscondB}
\eeq
for {\it all} $R\le R_0$. A sensible choice for the function ${\cal Y}$
is the following
\beq
{\cal Y}(R)=\left(\frac{1-\cos R}{1-\cos R_0}\right)^n,\;\;\;\;\;\;
n=1.
\label{Yfun}
\eeq
It has been proved in ref.~\cite{stefano} that such a choice allows
the definition of an isolated-photon-plus-jet cross section, which is
infrared-safe to all orders in QCD perturbation theory and still
does not receive any contribution from the fragmentation mechanism.
In this case, therefore, only the first term on the RHS of 
eq.~(\ref{factth}) is different from zero, and it does not contain any
$\mug$ dependence. The reader can find all the technical details
concerning the isolation prescription based on eq.~(\ref{iscondB})
of ref.~\cite{stefano}. Here, we will just recall the main ideas.
The fundamental property of the function ${\cal Y}$ is
\beq
\lim_{R\to 0} {\cal Y}(R)=0,
\label{limY}
\eeq
the function being different from zero everywhere except for $R=0$.
This implies that the energy of a parton falling into the isolation
cone ${\cal C}_{R_0}$ is correlated to its distance (in the $\eta$--$\phi$
plane) from the photon. In particular, a parton becoming collinear to
the photon is also becoming soft. When a quark is collinear to the photon,
there is a collinear divergence; however, if the quark is also soft,
this divergence is damped by the quark vanishing energy (provided that
the energy vanishes fast enough; this condition is not very restrictive,
and the form in eq.~(\ref{Yfun}) easily fulfils it). When a gluon is
collinear to the photon, then either it is emitted from a quark, which is
itself collinear to the photon -- in which case, what was said previously
applies -- or the matrix element is finite. Finally, it is clear that
the isolation condition given above does not destroy the cancellation
of soft singularities, since a gluon with small enough energy can be 
emitted anywhere inside the isolation cone. The fact that this prescription
is free of final-state QED collinear singularities implies that the 
direct part of the cross section is finite. As far as the fragmentation
contribution is concerned, in QCD the fragmentation mechanism is purely
collinear. Therefore, by imposing eq.~(\ref{iscondB}), one forces the
hadronic remnants collinear to the photon to have zero energy. This
is equivalent to saying that the fragmentation variable $z$ is restricted
to the range $z=1$. Since the parton-to-photon fragmentation functions
do not contain any $\delta(1-z)$, this means that the fragmentation
contribution to the cross section is zero, because an integration over
a zero-measure set is carried out. 

We stress that the function given in eq.~(\ref{Yfun}) is to a very large
extent arbitrary. Any sufficiently well-behaved function, fulfilling
eq.~(\ref{limY}), could do the job, the key point being the correlation
between the distance of a parton from the photon and the parton energy,
which must be strong enough to cancel the quark-to-photon
collinear singularity. We also remark that the traditional 
cone-isolation prescription, eq.~(\ref{iscondA}), can be recovered
from eq.~(\ref{iscondB}) by setting ${\cal Y}=1$ and $\epg=\epc$.
In the rest of this paper, as a short-hand notation, we will indicate
the `traditional' isolation obtained by imposing eq.~(\ref{iscondA}) as 
{\bf definition A}, and that obtained by imposing eq.~(\ref{iscondB}) as 
{\bf definition B}.

At first sight, the new isolation approach appears to be stricter than 
the traditional one. On the other hand, the fact that for the new constraint 
one also considers the angle between the photon and hadrons in the cone, 
is a real virtue here: for the traditional criterion, one would reject a 
hadron of, say, 2 GeV wherever it is located in the cone, just because its 
energy exceeds the limit. Of course, if the cone size 
is $0.7$, and the hadron has a distance of 0.6 with respect
to the photon -- why should one want to reject such an event?
This situation is improved with the new constraint: hadrons in the cone
that are still quite far away from the photon are allowed to have 
more energy than those close to the photon. In this way, one can well
allow a hadron to have 2 GeV, or even more, at a distance of 0.6. This 
little example implies that a detailed comparison between the traditional and 
the new isolation methods is certainly of some interest, and this will also be 
performed in this paper. One can then eventually decide which isolation to use 
in actual experiment.

\subsection{NLO computer codes}

In order to give phenomenological predictions, we will use two different
computer codes in the cases of definitions A and B. As far as definition
A is concerned, we use the NLO program of refs.~\cite{gv,gv2} for the 
direct part of the cross section. This program 
is based on an inclusive calculation of the contributing partonic
subprocesses $ab\to \gamma X$ to NLO, where $X$ contains a sum over the 
appropriate partonic final states, fully integrated over their phase
spaces. The advantage of this approach is that all contributions to
a given partonic channel, such as virtual and real-emission ones, and 
collinear counterterms, can be added before any numerical implementation.
In this way, not only all singularities that appear in the calculation cancel,
but there is no need to introduce any soft or collinear cut-off at 
intermediate stages of the numerical calculation either. One therefore 
ends up with a rather fast and accurate code. However, the drawback of this 
is the inclusiveness of the program: there is no handle on, say, an extra
jet since the partons in the final state have been integrated over.
As it stands, it would even seem impossible to implement isolation
in such a code since this clearly affords to have control over partons
falling into the isolation cone. However, as was shown in ref.~\cite{gv2},
one can fairly straightforwardly make up for this latter deficit, provided
the opening of the isolation cone is not too big. The idea is as 
follows: at NLO, the isolated cross section can also be viewed as the 
{\em non}-isolated one {\em minus} the cross section for a parton to be 
in the cone, having more energy than that allowed by the isolation cut.
This latter `subtraction' cross section can be approximated in a fairly
simple calculation in the limit of a narrow isolation cone, since it
is dominated by almost collinear quark--photon configurations. As a result,
the `subtraction' piece turns out to behave like ${\cal A} \ln R_0 + 
{\cal B} + {\cal O}(R_0^2)$; ${\cal A},{\cal B}$ are presented in~\cite{gv2}
for both the unpolarized and the polarized cases. Note that the `subtraction' 
piece itself inevitably depends on the final-state factorization scale 
$\mu_{\gamma}$ introduced in eq.~(\ref{factth}).

As we discussed in the previous subsection, for definition A the 
fragmentation mechanism contributes (cf. eq.~(\ref{factth})). 
We also stated in the introduction that in the polarized case we presently 
cannot calculate this part at NLO and thus have to stick to a LO calculation 
for it. In contrast to this, in the {\em un}polarized case the NLO corrections 
to the relevant partonic scatterings are known~\cite{aversa}. The calculation
and computer code presented in ref.~\cite{aversa} were also fully 
inclusive in the 
sense that the unobserved partons had been integrated over their full phase
spaces. However, as was shown in ref.~\cite{gv2}, the above `narrow-cone' 
approximation can also be used to implement isolation in the NLO 
fragmentation component as calculated with the program of ref.~\cite{aversa}. 
When calculating the unpolarized prompt-photon cross section for definition A
in this paper (for the purpose of computing asymmetries), we will always
include the fragmentation part at NLO, making use of the program 
of ref.~\cite{aversa}, along with the modifications for isolation developed
in ref.~\cite{gv2}.

The code relevant to definition B works in a completely different way,
being fully exclusive in the variables of the photon and of the (one or
two) final-state QCD partons. It is based on the formalism presented
in refs.~\cite{FKS,form97}, which allows the computation of any
infrared-safe observables for any kind of scattering particles,
without requiring algebraic manipulations on the matrix elements.
The formalism adopts the subtraction method in order to deal with
soft and collinear singularities, and therefore both the matrix
elements and the phase space are treated without any approximation.
The code used in this paper is an extension of that presented
in ref.~\cite{Vancouver}, which deals with the production of
isolated photons in {\it un}polarized hadronic collisions. Notice 
that the formalism of refs.~\cite{FKS,form97}, although originally
designed for the case of unpolarized collisions, extends -- basically
without any modifications -- to the case of polarized collisions.
A detailed discussion on this topic can be found in ref.~\cite{dfsv}.
We finally mention the fact that the code outputs the kinematical variables
of the photon and of the final-state partons, plus a suitable weight.
Therefore, the isolation condition, the jet-reconstruction algorithm,
and any cuts matching the experimental setup can be implemented as the 
final step of the computation algorithm. This allows us to plot as
many observables as we want in one single computer run.

It is easy to see that one can extend the ideas behind the `narrow-cone'
approximation, used for the definition-A code, also to the isolation
given by definition B. In this way, we have been able to compare
extensively the results of the two codes. We found excellent agreement 
of the two programs
over a wide range of kinematical variables, and also for cone openings
of even $R_0=0.7$, if only central values of rapidity are considered. 
This suggests the correctness of the two -- entirely
independent -- codes. It also implies that the `narrow-cone' approximation
has a rather large region of validity and can be well used for practical
applications. We recall, however, that the corresponding code is only
suitable for fully-inclusive photon observables, and not for
photon-plus-jet ones.

\section{Inclusive isolated-photon observables\label{sec:si}}

In this section, we study the inclusive properties of isolated photons.
More exclusive observables, such as correlations between the photon
and the accompanying jets, will be discussed in section~\ref{sec:corr}.
We will consider centre-of-mass
energies spanning the range \mbox{$\sqrt{S}=$200--500~GeV}.
We will carefully investigate the differences induced by the
different isolation prescriptions we deal with in this paper.
We will address the issue of the perturbative stability of our
results, and study the dependence of the cross sections upon the 
polarized parton densities. 

\subsection{Effects of isolation and discussion of theoretical 
uncertainties\label{subsec:si1}}
Unless otherwise specified, we 
will use the following parameters, as a default for our calculations:
\beqn
&&R_0=0.4,\;\;\;\;\epc=\frac{1~{\rm GeV}}{\ptg},
\;\;\;\;\;\;\;\;\;\;\;\;\;\;\;\;\!{\rm definition~A};
\label{isolparA}
\\
&&R_0=0.4,\;\;\;\;\epg=1,\;\;\;\;n=1,
\;\;\;\;\;\;\;\;\;\;{\rm definition~B}.
\label{isolparB}
\eeqn
It is worth emphasizing at this point that we have chosen $\epg \gg
\epc$: for traditional isolation A, $\epc$ has to be small --
otherwise, isolation is totally ineffective. For isolation B, on the 
other hand, $\epg$ may be chosen large, as we discussed previously.
A large $\epg$ only means that one still allows considerable amounts 
of hadronic energy in the cone, provided it is deposited far away from the
photon. We add that it is actually desirable
theoretically in any isolation to have a `large' value of $\epsilon$ ($=\epg$ 
or $\epc$): soft-gluon emission into the cone generates 
logarithms~\cite{bq1,gv2,stefano} of $\epsilon$, with an extra power at 
each further order of perturbation theory, which for very small 
$\epsilon$ eventually threaten to spoil the perturbative expansion. 
A study on the structure of the logarithms appearing in the 
isolated-photon cross section in $\epem$ collisions has been
given in ref.~\cite{CFP}. Pending a more thorough investigation 
of these points in the case of isolation B, it seems likely that 
being able to choose $\epg = {\cal O}(1)$ is clearly a virtue of 
this isolation method.

The default value for the factorization and renormalization scales
will be indicated by $\mu_0$, which will be taken equal to the
transverse momentum of the photon in the case A, and equal to half
of the total transverse energy of the event in the case B. These two choices,
which slightly differ beyond LO, are due to the different structure of the 
codes computing the isolated-photon cross section in cases A and B,
as described in section~\ref{sec:formalism}. It would be possible 
to set $\mu_0=\ptg$ in case B (however, this choice, although
formally correct, is less appropriate than the one adopted here:
since the code is fully exclusive in the variables of the photon 
and of the final-state partons, the reference scale, which is directly
related to the hardness of the process, should also depend upon the
transverse momenta of the partons); this would result in differences
with our default choice that are  completely negligible 
when compared to the other sources of theoretical uncertainty 
affecting the cross section. We will adopt throughout the two-loop
expression for $\as$, the $\Lambda_{\sss QCD}$ value being that
associated with the parton densities used. Our default parton
density sets will be the NLO `standard' set of ref.~\cite{GRSV} (GRSV STD) 
and MRST~\cite{MRST} for the polarized and unpolarized scattering 
respectively. In the case of the definition A, we will use the NLO
GRV~\cite{grvfrag} set of parton-to-photon fragmentation functions.

We have to note here that, while the value of $\Lambda_{\sss QCD}$ 
associated with the MRST set ($\lambdamsb=220$~MeV) is close to the 
central value of the latest PDG world average (at two loops,
$\lambdamsb=237_{-24}^{+26}$~MeV~\cite{PDG98}),
all the available polarized density sets have a value which is
much lower, consistent with that extracted from DIS data 
some years ago. Thus, by adopting the value of $\Lambda_{\sss QCD}$
associated with a given set, we have the unpleasant situation
in which, in the computation of asymmetries, the numerator and the
denominator have different $\Lambda$'s. Still, we preferred not
to violate the correlation between the parton densities and
$\Lambda_{\sss QCD}$. This correlation is expected to be particularly strong 
in the case of the gluon density, which is of great importance here.
Since a smaller $\Lambda_{\sss QCD}$ entails a smaller strong coupling,
our predictions for asymmetries would have become somewhat larger than the 
ones we present below, had we decided to adopt the same value of 
$\Lambda_{\sss QCD}$ in the polarized and unpolarized cross sections. This 
situation has already been encountered in ref.~\cite{dfsv}, for jet physics. 
There, it has been shown that using the same $\Lambda_{\sss QCD}$ in the 
polarized and unpolarized cross sections would increase the asymmetry by 
15\% (relatively) at the most. In the case of photon production, the 
difference is even smaller. As we will see, the effect is therefore 
completely negligible, with respect to the differences in the predictions of 
the asymmetries induced by the choice of different parton densities.

In what follows, in order to assess the importance of the radiative
QCD corrections, we will often compare the NLO and Born results.
Throughout the paper, by `Born result' we will mean the prediction
obtained by convoluting the lowest-order partonic cross sections
(\mbox{${\cal O}(\aem\as)$} and \mbox{${\cal O}(\as^2)$} for
the direct and the fragmentation contributions respectively) with
the NLO-evolved parton densities and, if needed, fragmentation functions.
Also, the two-loop expression
of $\as$ will be used. There is of course a lot of freedom in
the definition of a Born-level result. However, we believe that
with this definition one has a better understanding of some 
issues related to the stability of the perturbative series.
This is especially true in polarized physics, where the data
are not sufficient to determine the parton densities with a good accuracy, 
and where large (artificial) differences can arise between sets
fitted at LO or NLO to the available DIS data. For a detailed discussion on 
this point, see for example ref.~\cite{dfsv}.

\begin{figure}
\centerline{
   \epsfig{figure=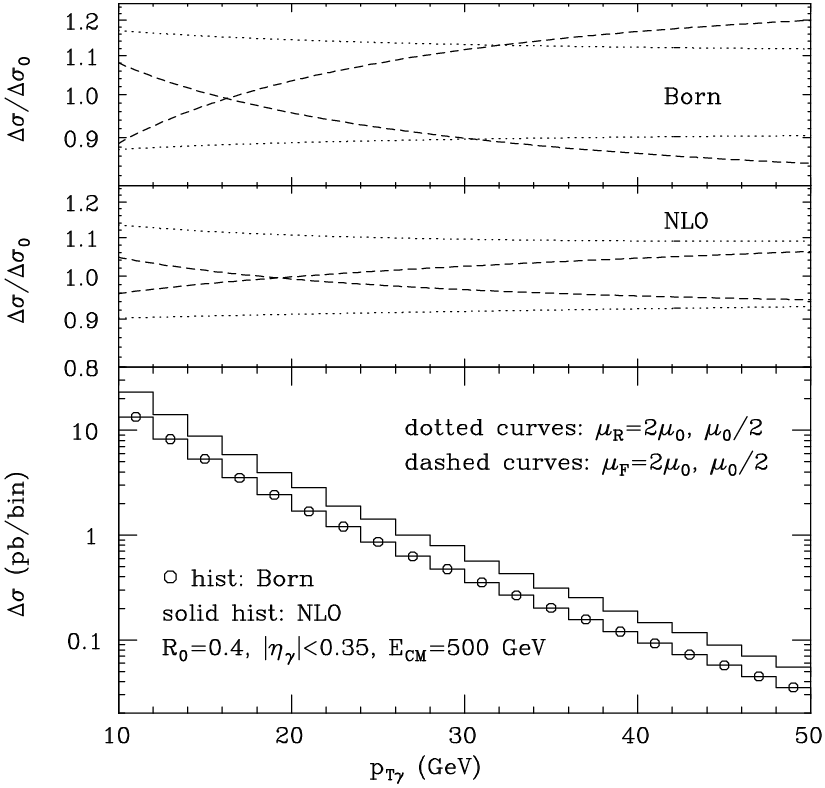,width=0.7\textwidth,clip=}
           }
\ccaption{}{ \label{fig:ptspectrumB}
Transverse-momentum spectrum of the isolated photon, in the case
of definition B, for polarized $pp$ collisions at $\sqrt{S}=500$~GeV.
The polarized parton densities used are GRSV STD.
The scale dependences of the Born and NLO results
are also shown; see the text for details.
}
\end{figure}                                                              
We start by considering the transverse momentum spectrum of isolated
photons. In the lower part of fig.~\ref{fig:ptspectrumB}
we plot the Born (histogram with symbols) and NLO (solid histogram)
results for the polarized cross section, obtained at $\sqrt{S}=500$~GeV 
with the isolation definition B. A cut \mbox{$\abs{\etag}<0.35$}
has been imposed, which is suitable for the PHENIX experiment. 
As can be seen from the figure, the
inclusion of the radiative corrections gives a sizeable effect
as far as the normalization is concerned (in the first bin,
the ratio of NLO over Born result is about 1.8), while the
shape is almost unaffected (the Born being only slightly harder
than the NLO result). Since the radiative corrections are large,
one may wonder whether the NLO result is a sensible quantity
to compare with data. A rigorous answer to this question can
only come from a complete NNLO calculation. Lacking that, we
study the scale dependence of our results, as customary in
perturbative QCD, to see whether the inclusion of radiative 
corrections leads to a milder dependence upon the scales, as
compared to the one of the Born result. Here, it is especially
important to study the separate dependence upon the renormalization
and factorization scales, because cancellation effects between
the two may hide some problems. We present the scale dependence
of the $\ptg$ spectrum in the upper and central parts of 
fig.~\ref{fig:ptspectrumB}. There, we show the ratio of the cross 
section obtained by setting the scales equal to $\mu_0/2$ and $2\mu_0$, 
over the cross section for the default values of the scales. 
We stress that only one scale is varied at a time. The renormalization
(factorization) scale variation corresponds to the dotted (dashed)
curves; the curves decreasing for increasing $\ptg$ correspond to 
$\mur=\mu_0/2$ and $\muf=2\mu_0$, respectively. From the figure, 
it is apparent that the inclusion of the
radiative corrections reduces the scale dependence in the whole
$\ptg$ range considered, with a possible exception in the case
of the $\muf$ dependence, for $\ptg$ equal to 16--20~GeV, 
where there is basically no $\muf$ dependence. The reduction 
is stronger in the case of $\muf$ dependence than in the 
case of $\mur$ dependence. The fact that there is a point in the 
$\ptg$ spectrum where there appears to be no factorization scale dependence
is purely accidental; it can be traced back to the behaviour of
the parton densities with respect to the hard scale. In fact, 
the gluon density increases with the scale 
in the $x$ range corresponding to the low-$\ptg$ region, while
it is decreasing when the scale is increasing for larger $x$ values, 
probed when a harder photon is produced. We can conclude from 
fig.~\ref{fig:ptspectrumB} that the perturbative expansion
seems to be reliable in this case; in all cases, the radiative 
corrections reduce the size of the dependence of the $\ptg$ 
spectrum upon the scales. We must comment on the fact that this
conclusion is not specific to the kinematical configuration 
considered in fig.~\ref{fig:ptspectrumB}: we verified that
the same kind of behaviour can be seen in a larger $\etag$ 
range (we studied the case \mbox{$-1<\etag<2$}), and also
at lower centre-of-mass energies ($\sqrt{S}=200$~GeV).
Furthermore, almost the same results are obtained
in the case of unpolarized collisions.

\begin{figure}
\centerline{
   \epsfig{figure=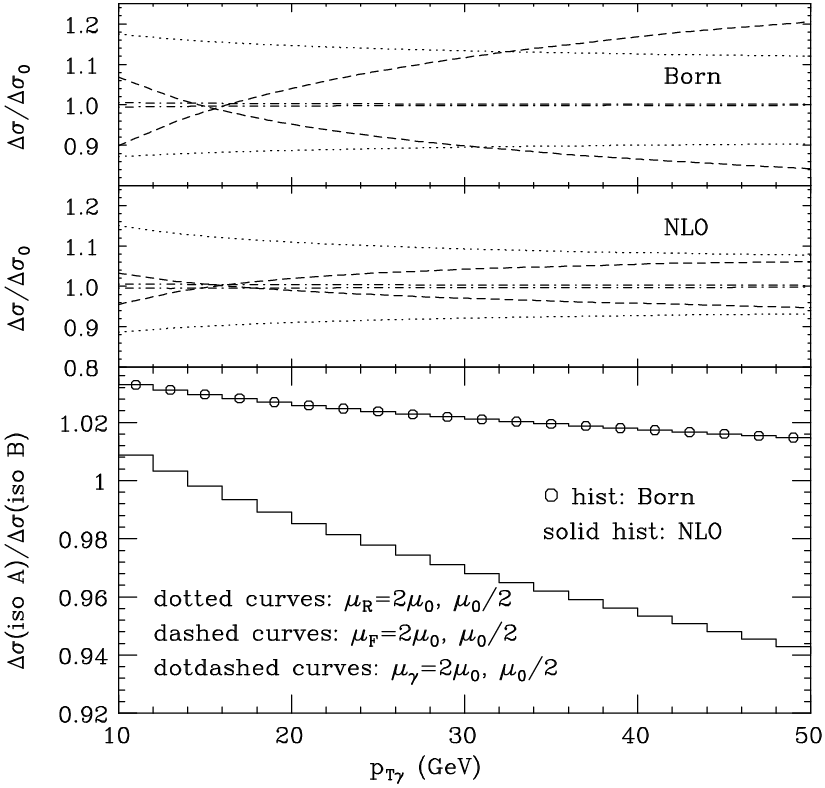,width=0.7\textwidth,clip=}
           }
\ccaption{}{ \label{fig:ptspectrumA}
Transverse-momentum spectrum of the isolated photon, in the case
of definition A. The ratio of the cross section over that obtained
with definition B is shown in the lower part. The rest of the figure
displays the scale dependence, at the Born and NLO levels.
}
\end{figure}

We now turn to the case when the photon is isolated according 
to definition A. The results are presented
in fig.~\ref{fig:ptspectrumA}. In the lower part, we display the
ratio of the cross section over that obtained with definition
B. In this case, the scales are fixed to their default values.
The result at the Born level is again displayed as a histogram
with symbols. The Born result in the case of definition A is
always higher than that relevant to definition B. This is 
easy to understand, since at this order the result for the 
direct part is independent of the isolation condition, and 
the photon isolated with definition A gets a contribution
from the fragmentation part, which is not present in the case
of definition B. Things of course change at NLO: having an
additional parton around, the isolation condition is effective also
in the direct part. We must also remark that, in the case of isolation A,
the fragmentation contribution is only included at LO. A consistent
computation at NLO would presumably produce a slightly larger cross 
section (for example, in the case of unpolarized collisions, the inclusion
of radiative corrections in the fragmentation component enhances
the full cross section at high $\pt$ by about 3\%). The effect is
much larger in the case in which there is no isolation condition,
and the photon is fully inclusive. We will further comment on this 
fact below. As in the previous case, we also studied the $\mur$
and $\muf$ dependence of the spectrum; in doing so, the factorization 
and renormalization scales of the direct and of the fragmentation components
have been set to the same value: $\mufp=\muf$ and $\murp=\mur$
(see eq.~(\ref{factth})). The results are displayed in the
upper parts of fig.~\ref{fig:ptspectrumA}. Note that for definition A
we have an additional pair of lines (dot-dashed), corresponding
to the results obtained by varying the {\em final}-state factorization
scale $\mug$, which enters the fragmentation functions.
The $\muf$ and $\mur$ dependence is very similar to the one
relevant to definition B, displayed in fig.~\ref{fig:ptspectrumB},
and the same comments made previously apply here. On the other
hand, the $\mug$ dependence is extremely small, and gives a negligible
contribution to the theoretical error affecting the cross section.
The almost identical scale dependence in the case of definitions A and B
also implies that the ratio of cross sections plotted in the lower
part of fig.~\ref{fig:ptspectrumB} is, to a good extent, independent
of the scale choice.

\begin{figure}
\centerline{
   \epsfig{figure=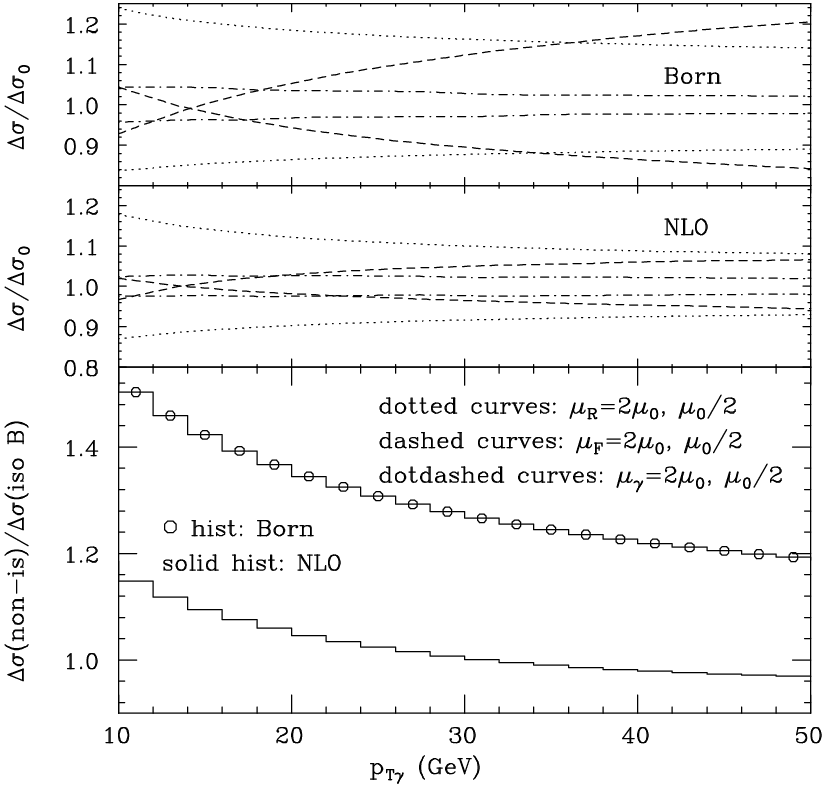,width=0.7\textwidth,clip=}
           }
\ccaption{}{ \label{fig:ptincl}
Same as in fig.~\ref{fig:ptspectrumA}, for non-isolated photons.
}
\end{figure}

For completeness, we present in fig.~\ref{fig:ptincl} the
corresponding predictions for the fully inclusive {\it non-isolated}
prompt-photon cross section. We first discuss the scale dependence of
the results, displayed again in the upper two parts of the figure. At
low $\ptg$, the $\mur$ dependence turns out to be larger than that of
the isolated-photon cross sections, while at large $\ptg$ the two
appear to be pretty similar (the isolation condition is less and less
restrictive as the transverse momentum of the photon is increased,
since it is more and more difficult to have a hard parton, in the
surroundings of the photon, that does not pass the isolation cuts).
The $\mug$ dependence of the fully inclusive cross section is much
larger than that of the isolated-photon cross section obtained with
definition A. However, its effect is still smaller than that due to
$\mur$ and $\muf$.

The comparison of the non-isolated cross section with the one isolated 
according to definition B is shown in the lower part of fig.~\ref{fig:ptincl}.
One immediately notices a striking feature: at NLO, the isolated cross 
section becomes larger than the unpolarized one at large $\ptg$. 
Clearly, this finding is at odds with the physical expectation that any 
meaningful isolation cut should reduce the number of events with respect to 
the number obtained for no isolation at all. 
The origin of the problem we encounter 
resides in the fact that the fragmentation contribution to the non-isolated 
cross section (and, obviously, also for the results for isolation A
presented in fig.~\ref{fig:ptspectrumA}) has only been calculated at the
LO level since, as we pointed out earlier, the NLO corrections to
the fragmentation contribution have not been calculated so far
in the polarized case. We expect that once the proper NLO fragmentation
component is included in the calculation of the polarized cross
section, the disagreeable feature of fig.~\ref{fig:ptincl} will 
disappear. This view is corroborated by the observation that
we find exactly the same pattern in the unpolarized case: 
there, everything can be calculated consistently at NLO, and
the non-isolated cross section turns out to be larger than the one obtained 
for both types of isolation we consider. However, we checked that, 
{\em if} we compute the fragmentation contribution only at LO 
in the unpolarized case, we indeed obtain a 
non-isolated cross section that is {\em smaller} than the isolated one
for definition B, much as happens in fig.~\ref{fig:ptincl}.
The figure clearly points out the importance of consistency in the 
NLO calculation. A NLO calculation of the fragmentation component of the 
polarized prompt-photon cross section is highly desirable for the 
future. On the basis of fig.~\ref{fig:ptincl} we would predict non-negligible
positive corrections to the LO result. For the time being, our present 
results for isolation A and the non-isolated case have only limited 
reliability. Fortunately, fragmentation is really important only in
the non-isolated case, which is not the one relevant to experiment.
For a (traditionally) isolated cross section, it contributes a relatively 
small fraction of the full result, and this fraction decreases rapidly 
towards larger $\ptg$. We are therefore fairly confident that our 
predictions for definition A are numerically not too far off the true 
NLO answer.

For these reasons, we refrain from performing a detailed study of the 
uncertainty in the cross section for isolation A. We 
only state that we have also calculated the unpolarized and the 
polarized cross sections, using set I of the photon fragmentation
functions of ref.~\cite{bfg}. We find that the fragmentation component
to the cross section decreases by about 23\% at $\ptg\approx 
10$~GeV, and by about 50\% at the high-$\ptg$ end. For the full (i.e.
direct plus fragmentation) cross section, the effect is obviously 
much smaller, generally below 2\%.

\begin{figure}
\centerline{
   \epsfig{figure=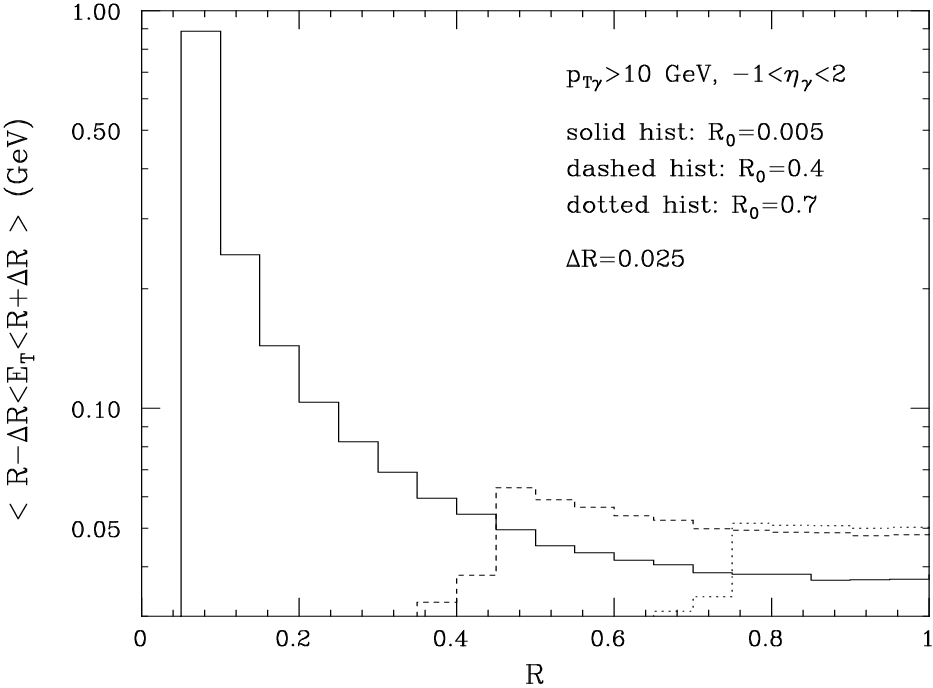,width=0.7\textwidth,clip=}
           }
\ccaption{}{ \label{fig:conedep}
Average transverse energy in the cone annuli around the photon.
}
\end{figure}

There is another striking property of the curve in the lower part 
of fig.~\ref{fig:ptspectrumA}: it is {\em very} close to unity over the 
whole range of $\ptg$ (in fact, from what we just discussed, we would
expect it to be even closer to unity at large $\ptg$, had we been able to 
include the fragmentation component at NLO in case of definition A).
In other words, the two types of isolation, albeit so different 
from a physics point of view, lead to almost identical cross sections.
To some extent, this is certainly due to the choices we made for
$\epc$, $\epg$ in eqs.~(\ref{isolparA}),(\ref{isolparB}): had we 
chosen, say, $\epg=\epc$ there\footnote{This is simply an example; as 
discussed at the beginning of this subsection, this is actually not a 
desirable choice.}, then isolation B would have become
stricter than isolation A, and the corresponding curve in 
fig.~\ref{fig:ptspectrumA} would have been above unity everywhere.
Our choices in eqs.~(\ref{isolparA}),(\ref{isolparB}) presumably
created a certain `balance' between the two isolations.
However, we found that there is more to the similarity of the two
isolated cross sections. When performing runs at larger values
of $R_0$, we found that the cone-size dependence of the cross
section is extremely mild for both types of isolation. This indicates
that isolation is most effective close to the photon and does not
affect the cross section too much at larger distances from the photon.
To illustrate this point, we plot in fig.~\ref{fig:conedep}, as a 
function of the distance $R$ from the photon, the amount of hadronic 
transverse energy deposited on average in a cone annulus between 
$R-\Delta R$ and $R+\Delta R$, where $\Delta R=0.025$. We do this for 
isolation of type B, considering two realistic values of the isolation-cone 
size, $R_0=0.4$ and $R_0=0.7$, and one extreme value $R_0=0.005$. 
We have chosen $\sqrt{S}=500$~GeV, and the photon variables 
have been integrated over $\ptg>$ 10 GeV, $-1<\etag<2$ (we used an
extended rapidity coverage in order to reduce as much as possible
the statistical errors on our results; however, rapidity is not
an issue here, and our conclusions apply to any rapidity range).
The reason for having a plot obtained with $R_0=0.005$ is the following:
with such a narrow isolation cone, the distribution in hadronic transverse 
energy around the photon is basically identical (for non-zero $R$) to what
would be obtained in the case of non-isolated photons. Thus, we can have
a clear idea of the effect of imposing an isolation condition with
a realistic value for $R_0$. By inspection of the solid histogram in 
fig.~\ref{fig:conedep}, we see that a sizeable amount of energy is deposited 
only at small distances from the photon, and -- at least in the framework 
of a NLO calculation -- not much energy is delivered for 
\mbox{$R>$0.2--0.3}. This is the reason why there is no real 
difference between the cross sections obtained with different,
but physically sensible, values of $R_0$. Notice that this conclusion
holds regardless of whether we use isolation A or B. 
To avoid misunderstandings concerning 
fig.~\ref{fig:conedep}: of course at NLO we have only one extra particle
at our disposal (the third one balancing the other two in transverse 
momentum), so isolation only becomes effective if this particle falls into the
isolation cone. We are not saying in fig.~\ref{fig:conedep} that this extra 
particle is usually soft: the point is rather that, thanks to the 
collinear singularity of the corresponding $2\to 3$ matrix elements 
at $R=0$, the extra particle simply happens to be close to the 
photon more often than far away from it, so that {\em on average}
more energy is deposited close to the photon. Incidentally, one can 
convince oneself that, for small $R$ and $\Delta R \ll R$, 
the quantity shown in fig.~\ref{fig:conedep} has to be proportional to 
$\Delta R/\sigma(R_0) \times ( d\sigma (\theta)/d\theta)_{\theta=R}$, 
where $\theta$ is the angle between the momenta of the photon and the  
other particle in the cone, and $\sigma(R_0)$ is the total cross
section for a given isolation-cone size $R_0$ and a given kinematical
range for the photon variables (here, $\ptg>10$~GeV and $-1<\etag<2$).
For a quark parallel to the photon (which is the only configuration
producing a collinear singularity at $\theta=0$), one thus finds
immediately that the curves in fig.~\ref{fig:conedep} should fall
$\propto 1/R$ if $R>R_0$, i.e. outside the isolation cone. This is
exactly the pattern observed in the figure.  Inside the cone the
isolation is effective, and for the isolation of definition B employed
here, one expects the curve to fall like some power of $R$ as $R\to
0$. Finally, we also note that the normalization factor
$1/\sigma(R_0)$ is the reason why the histograms for the three
different $R_0$ in fig.~\ref{fig:conedep} do not exactly coincide even
if $R>0.7$. A comparison of the histograms relevant to $R_0=0.4$ and
$R_0=0.7$ at $R=1$ nicely demonstrates how weak the dependence of the
isolated-photon cross section on the cone size is.

We finally study the distribution of the photon in pseudorapidity $\etag$.
We impose here 
\begin{figure}[ht]
\centerline{
   \epsfig{figure=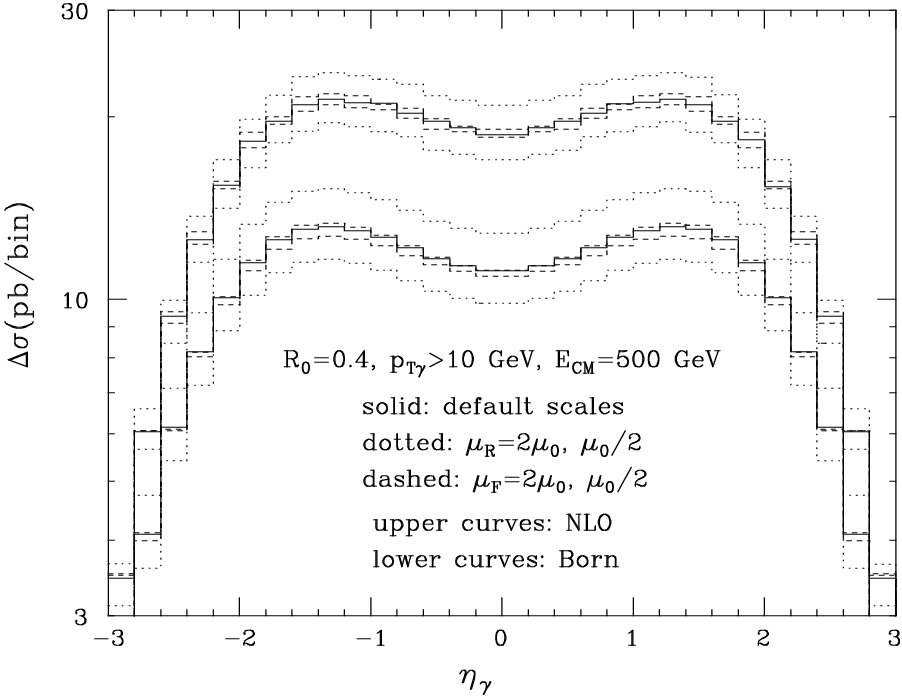,width=0.7\textwidth,clip=}
           }
\ccaption{}{ \label{fig:etaB}
Pseudorapidity spectrum of isolated photons (definition B).
The cross sections obtained by varying the renormalization
and factorization scales are also shown.
}
\end{figure}                                                              
\noindent
the transverse momentum cut $\ptg>10$~GeV. Our
results are summarized in fig.~\ref{fig:etaB}; they have been
obtained in the case of definition B. The upper histograms 
correspond to NLO, while the lower ones represent the Born level
results. The solid histograms have been obtained with default scales.
They are pretty similar in shape; the Born is only slightly broader,
with a deeper dip at $\etag=0$. The size of radiative corrections
is as expected from what we previously found for the $\ptg$ spectrum:
the NLO result is about a factor 1.6 higher than the Born one.
The pairs of dotted histograms are obtained by setting the
renormalization scale equal to $\mu_0/2$ and $2\mu_0$. It is
clear that also in this case the radiative corrections have the
effect of reducing the scale dependence; by varying $\mur$,
one obtains NLO cross sections that differ from the default one
by 10\% at most. The effect of varying the factorization
scale is much smaller than that associated with changes in $\mur$. This
can be understood by looking at fig.~\ref{fig:ptspectrumB}:
the effects at low $\ptg$ ($\muf=\mu_0/2$ returns a cross
section smaller than the default one) and at large $\ptg$
($\muf=\mu_0/2$ returns a cross section larger than the default 
one) tend to compensate when integrating over $\ptg$, as done for
fig.~\ref{fig:etaB}. The scale dependence
of the densities is again responsible for this behaviour.
When plotting the $\etag$ distribution for an integration over, say,
just the region \mbox{$10<\ptg<12$~GeV}, one would see a much larger $\muf$
dependence, of the order of that shown in the low-$\ptg$ region
in fig.~\ref{fig:ptspectrumB}. We also considered the $\etag$ spectrum
in the case when the photon is isolated according to definition A.
The scale dependence, as already in the case of the $\ptg$ spectrum,
is identical to that presented in fig.~\ref{fig:etaB}. The shape
is almost the same as the one relevant to definition B, for
all scale choices.

\subsection{Spin asymmetries and sensitivity to $\Delta g$\label{subsec:si2}}
In spin physics, experiments usually focus on spin asymmetries,
since many systematic uncertainties cancel out 
in this ratio of polarized and unpolarized cross sections. 
In what follows, we will therefore study the quantity
\beq
{\cal A}_{\ptg}=\frac{d\Delta\sigma/d\ptg}{d\sigma/d\ptg},
\label{asypt}
\eeq
as a function of $\ptg$. We will also consider a similar asymmetry,
with $\ptg$ replaced by $\etag$. More studies on asymmetries will
be presented in section~\ref{sec:corr}. In eq.~(\ref{asypt}), it
is understood that the same kinematical cuts are applied to both
the numerator and the denominator. The measurability of a spin asymmetry
for a given process, as far as statistics are concerned, is of course 
determined by the counting rate. The quantity 
\beq
\left({\cal A}_{\ptg}\right)_{min}=\frac{1}{P^2}
\frac{1}{\sqrt{2\sigma {\cal L}\ep}}
\label{minasy}
\eeq
can be regarded as the minimal asymmetry that can be detected experimentally
or, equivalently, as the expected statistical error of the measurement,
for a given integrated luminosity relevant to parallel or antiparallel
spins of the incoming particles, ${\cal L}$, 
beam polarizations $P$ and a detection efficiency $\ep\le 1$;
$\sigma$ is the unpolarized cross section integrated over a certain range 
in $\ptg$ (this range is denoted as a $\ptg$ bin). 
We assumed here that the luminosities relevant to parallel and antiparallel
spins of the incoming protons are equal, \mbox{${\cal L}^{\rightrightarrows}=
{\cal L}^{\rightleftarrows}\equiv{\cal L}$}. If this were not the case,
the quantity \mbox{$2\sigma {\cal L}$} in eq.~(\ref{minasy}) would have
to be substituted with
\mbox{$4\sigma {\cal L}^{\rightrightarrows}{\cal L}^{\rightleftarrows}/
({\cal L}^{\rightrightarrows}+{\cal L}^{\rightleftarrows})$}.

At present, the largest source of uncertainty on the theoretical
predictions for asymmetries is clearly the choice of the polarized parton
densities. We established in the previous subsection 
that the perturbative expansion
of single-inclusive cross sections is under control. In the remainder
of this section, we will therefore concentrate on the dependence of
the asymmetries upon the densities. There are many NLO-evolved
sets available, which mainly differ in the gluon sector, which is
rather poorly constrained by DIS data (apart, perhaps, from the first moment).
In a previous study relevant to jet physics~\cite{dfsv}, we 
saw that most of the sets give almost identical results as far
as the shape of the asymmetry is concerned, the main difference
being in the absolute normalization. For this reason, we will limit ourselves 
in this paper to three sets: GRSV STD, our default
set; GRSV MAXg~\cite{GRSV}, which has a much larger gluon density and is thus 
expected to return the highest cross sections; and set C of ref.~\cite{GSC}
(GS-C), which has a rather small gluon density whose shape is dramatically
different from that of all the other sets, turning negative at high $x$
for $Q^2$ not too large.

\begin{figure}
\centerline{
   \epsfig{figure=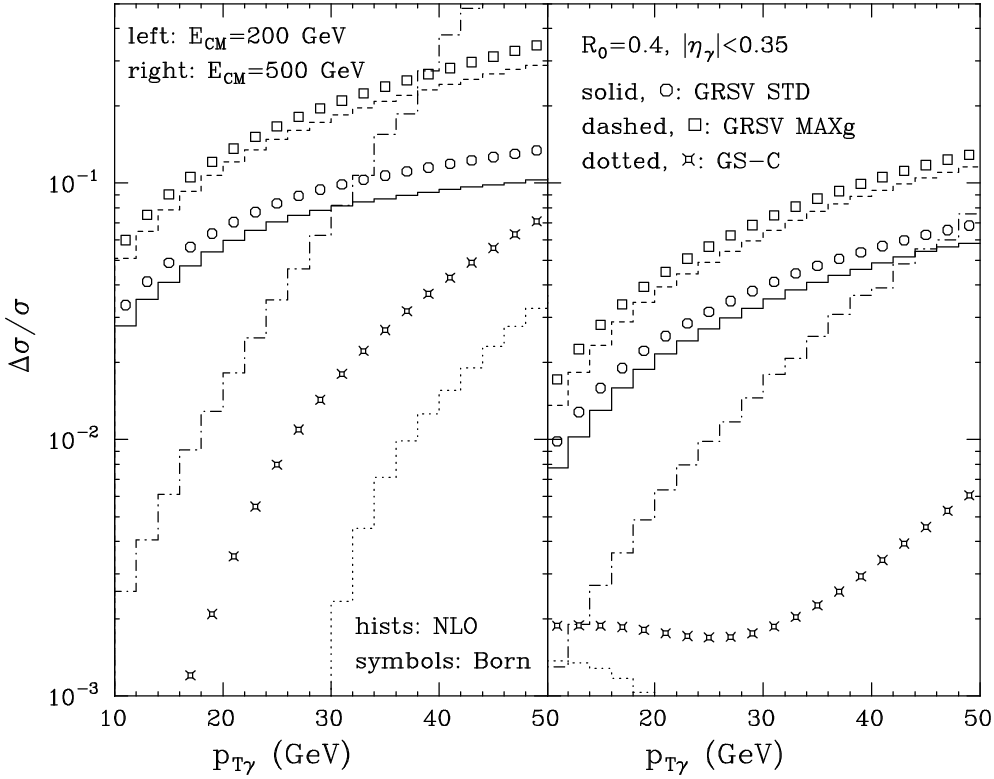,width=0.7\textwidth,clip=}
           }
\ccaption{}{ \label{fig:asypt}
Asymmetry as a function of $\ptg$, for various polarized densities,
at different centre-of-mass energies. The minimally observable 
asymmetry (dot-dashed histogram) is also shown.
}
\end{figure}

In fig.~\ref{fig:asypt} we present our results for the asymmetry 
as a function of $\ptg$. A cut \mbox{$\abs{\etag}<0.35$} has
been applied. In the left part of the figure we plot the asymmetries
obtained at $\sqrt{S}=200$~GeV, while in the right part we present
the results for $\sqrt{S}=500$~GeV. The solid, dashed and dotted
histograms correspond to the NLO predictions obtained with 
GRSV STD, GRSV MAXg and GS-C respectively. The corresponding
symbols (see the labels on the figure) are the results obtained
at the Born level. Finally, the dot-dashed histogram is the
minimally observable asymmetry, as defined in eq.~(\ref{minasy}).
We have chosen ${\cal L}=100$~pb$^{-1}$, $P=1$ and $\ep=1$.
Of course, the latter two choices are not realistic; however,
in adopting this `ideal-world' situation, we can estimate the
optimally achievable accuracy for a given integrated luminosity.
Note that the assumed value for ${\cal L}$ is conservative;
one expects in the best case to eventually obtain 
${\cal L}=160$~pb$^{-1}$/polarization at $\sqrt{S}=200$~GeV and
${\cal L}=400$~pb$^{-1}$/polarization at $\sqrt{S}=500$~GeV. In any 
case, it is straightforward to rescale our predicted minimally 
observable asymmetry~\footnote{Our cuts used in fig.~\ref{fig:asypt}, 
in particular $\abs{\etag}<0.35$, actually correspond to those of the 
PHENIX experiment. Note, however, that all our results have been 
integrated over the full $2\pi$ of azimuthal angle, whereas the 
PHENIX electromagnetic calorimeter only covers half the azimuth.
This implies that for a correct comparison our minimally observable 
asymmetry has to be multiplied by a factor of $\sqrt{2}$, in addition
to introducing the appropriate values for ${\cal L}$, $P$, $\ep$.
Our results for $\left({\cal A}_{\ptg}\right)_{min}$ are then found 
to be consistent with those reported in~\cite{goto}.} 
if one prefers other values for 
${\cal L}$, $P$ or $\ep$. We also emphasize that we have chosen
rather small bins in $\ptg$, $\Delta \ptg=2$~GeV. It would 
certainly seem advantageous in the actual data analysis to 
increase the bin size when going to larger $\ptg$, as is indeed
a commonly adopted procedure in the unpolarized prompt-photon experiments. 

From fig.~\ref{fig:asypt}, we see that the shapes of the asymmetries
obtained using the GRSV STD and GRSV MAXg sets are quite similar, but the
difference in normalization is sizeable; this is consistent with
what we observed in our study of 
jet physics~\cite{dfsv}. On the other hand, the 
result obtained for GS-C looks completely different. At NLO,
this asymmetry becomes negative in the region \mbox{$10<\ptg<30$~GeV}
at $\sqrt{S}=200$~GeV, and in the region \mbox{$30<\ptg<45$~GeV}
at $\sqrt{S}=500$~GeV. This is due in the first place to large
cancellations between the contributions of various partonic channels:
while the GRSV STD and GRSV MAXg results are dominated by the contribution
of the $qg$-initiated subprocess, this is not true in the case of GS-C, 
where the gluon is so small that quark--quark scatterings (of opposite
sign) are in absolute value of the same order or larger, in particular 
in the central pseudorapidity region, which is of interest here.
For the same reason, the asymmetries for set GS-C obtained at the Born
level turn out to be always substantially larger than those obtained at NLO.
The issue of (non-)dominance of the $qg$ subprocess, which is obviously 
of key interest for the extraction of the polarized gluon density from 
isolated-photon data, will be examined in more detail in the following.

It is instructive to compare the asymmetries at the two different 
centre-of-mass energies considered in fig.~\ref{fig:asypt}. As is well known, 
at smaller centre-of-mass energies, the asymmetries are generally larger; 
for example, when going from $\sqrt{S}=200$~GeV to $\sqrt{S}=500$~GeV
the asymmetry obtained with GRSV STD decreases
by a factor of about 3.6 (1.8) at $\ptg=10$~GeV ($\ptg=50$~GeV).
This feature is readily explained by the fact that, for fixed 
$\ptg$, at larger $\sqrt{S}$ one probes smaller values of $x$ 
in the parton distributions. Since the unpolarized parton densities are 
steeper than the polarized ones towards small $x$'s, one therefore gives 
more weight to the unpolarized cross section in the denominator of the 
asymmetry. However, when comparing the predicted asymmetries with the minimum 
observable asymmetry, it is clear that, at a fixed value of $\ptg$,
and except for the first few $\ptg$ bins, the situation at $\sqrt{S}=500$~GeV
is more favourable than that at $\sqrt{S}=200$~GeV.
On the other hand, as far as a measurement of $\Delta g$ at a given $x$ is 
concerned, one should rather look at the asymmetries at fixed 
$\xtg=2 \ptg/\sqrt{S}$, since this corresponds to the value at which 
the parton densities are probed predominantly. As can be seen from
fig.~\ref{fig:asypt}, the asymmetries corresponding to the GRSV sets
approximately scale with $\xtg$. 
Then, the quantity deciding about which energy is more favourable, is the 
minimally observable asymmetry at a given $\xtg$. This quantity does {\em not}
scale with $\xtg$, as can be inferred from the figure. For $\sqrt{S}=500$~GeV,
one finds a value of $\left({\cal A}_{\ptg}\right)_{min}$ larger than for 
the lower energy, making the higher-energy option appear less favourable. 
However, two points should be kept in mind here: firstly, in both plots
in fig.~\ref{fig:asypt} the same value for the integrated luminosity
has been used, whereas in reality one anticipates a higher (by a factor 
of 2 to 3) luminosity for $\sqrt{S}=500$~GeV. Secondly, the {\em lower} 
cut-off for $\ptg$ will certainly be the same for both energies, 
which means that at $\sqrt{S}=500$~GeV one can explore a region of
$\xtg$ that is inaccessible at $\sqrt{S}=200$~GeV. Also, even if one
considers the same $\xtg$ value for the two energies,
the parton densities are still being probed at rather different scales,
of the order of \mbox{$\xtg\sqrt{S}/2$}, the corresponding $\ptg$ values.  
Thus, it will be interesting to see whether measurements performed at 
different centre-of-mass energies will yield information that is consistent, 
and compatible with QCD evolution.

The same asymmetries as presented in fig.~\ref{fig:asypt} were 
also computed in a larger pseudorapidity range, \mbox{$-1<\etag<2$},
relevant to the STAR experiment.
In the case of the GRSV sets, only very minor differences were
found, and the same conclusions as drawn above apply (we have also to
take into account the fact that the minimally observable asymmetry 
decreases by a factor that can be as large as 2; therefore, in this larger 
pseudorapidity range the situation is even more favourable). In the case of
GS-C, the asymmetry increases significantly, becoming larger than the
minimally observable asymmetry for $\ptg<22$~GeV at $\sqrt{S}=500$~GeV,
and for $\ptg<14$~GeV at $\sqrt{S}=200$~GeV. Also, in this region
of $\ptg$ the Born and NLO results are close to each other, displaying
a behaviour similar to that of the GRSV sets. The reason is again
linked to the dominance of the $qg$-initiated partonic subprocess:
in fact, at large $\etag$'s, this subprocess accounts for most of 
the full NLO cross section also in the case of GS-C, as will
be shown below. It thus follows that,
if the true polarized gluon density is similar to the one of the GS-C
set, an extended pseudorapidity coverage is mandatory in order to
be able to see it experimentally. Finally, we mention 
that we computed the asymmetries also by varying the isolation
parameters $R_0$ and $n$. When setting $R_0=0.7$ and/or $n=2$,
we did not find any noticeable difference with the results 
presented above.

\begin{figure}
\centerline{
   \epsfig{figure=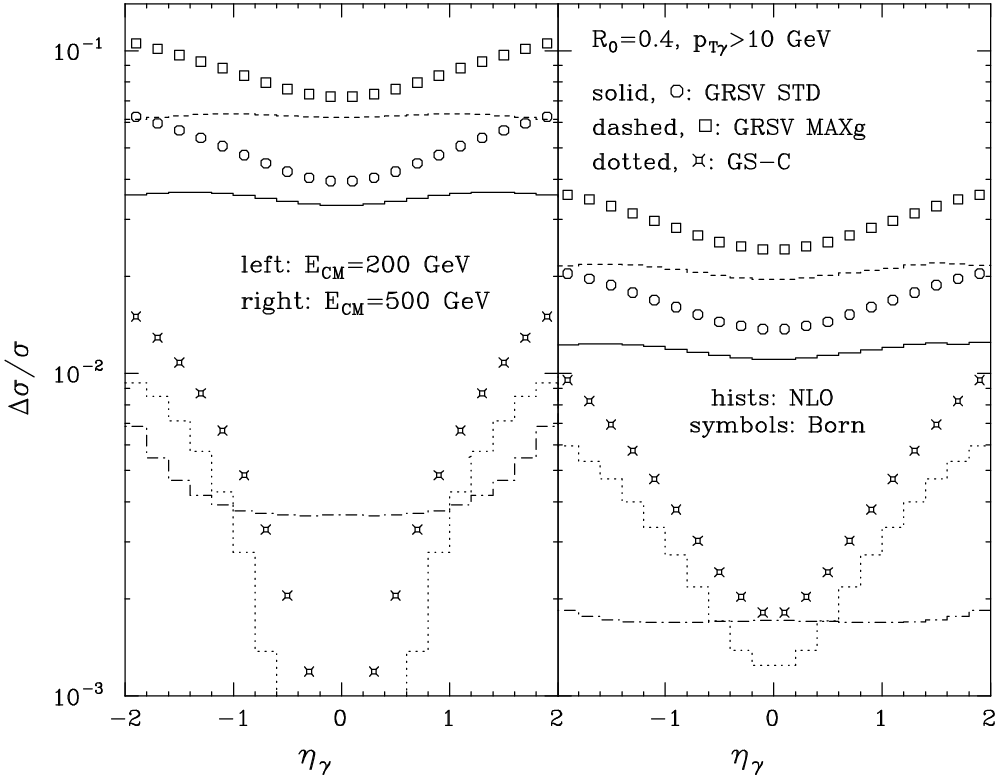,width=0.7\textwidth,clip=}
           }
\ccaption{}{ \label{fig:asyeta}
As in fig.~\ref{fig:asypt}, but as a function of $\etag$.
}
\end{figure}                                                              
In fig.~\ref{fig:asyeta} we present the asymmetry as a function
of $\etag$. A cut $\ptg>10$~GeV has been applied. We again show
the results obtained at two different centre-of-mass energies,
at both the Born and the NLO level. As in the case of the $\ptg$
distribution, the shapes of the asymmetries obtained with the two 
GRSV sets are almost identical, while the one obtained with GS-C 
behaves quite differently. The NLO result for GS-C becomes negative 
around $\etag=0$ at $\sqrt{S}=200$~GeV (the Born asymmetry remains 
positive), consistently with what we observed before in the low-$\ptg$ 
region at this energy. As remarked before, away from the central-$\etag$ 
region the GS-C asymmetry is larger than the minimally observable one,
and the Born and the NLO results become similar. On the other hand, 
in this region the Born and NLO results for the GRSV sets differ more 
than around $\etag=0$. This is simply related to the fact that here 
the polarized cross sections fall more rapidly at the NLO level
than at the Born level (cf. fig.~\ref{fig:etaB}), contrary to the 
case of the unpolarized cross section.

\begin{figure}
\centerline{
   \epsfig{figure=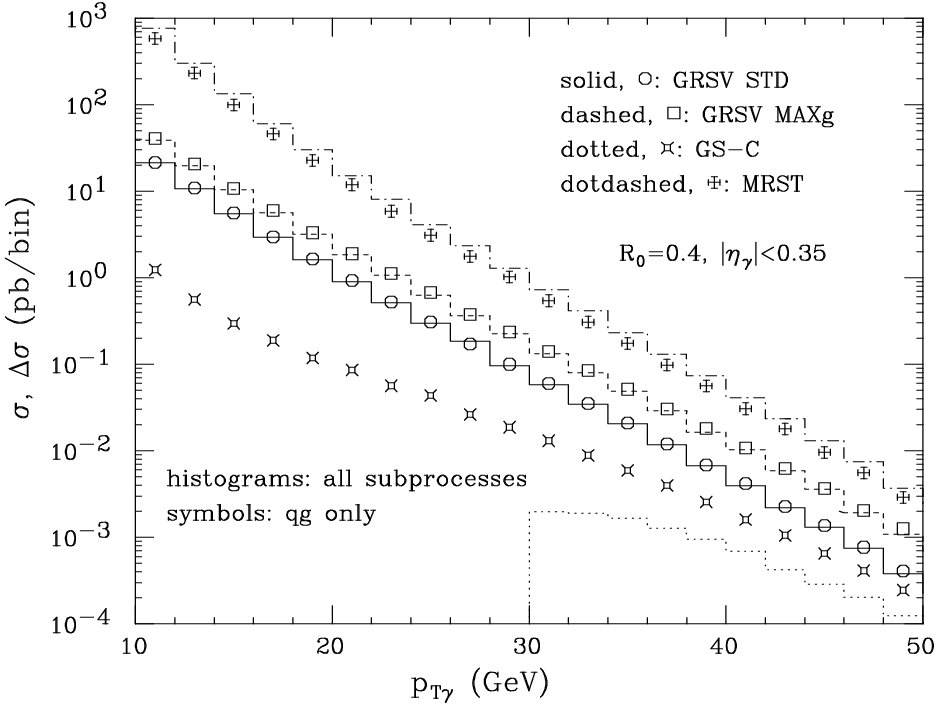,width=0.48\textwidth,clip=}
   \hfill
   \epsfig{figure=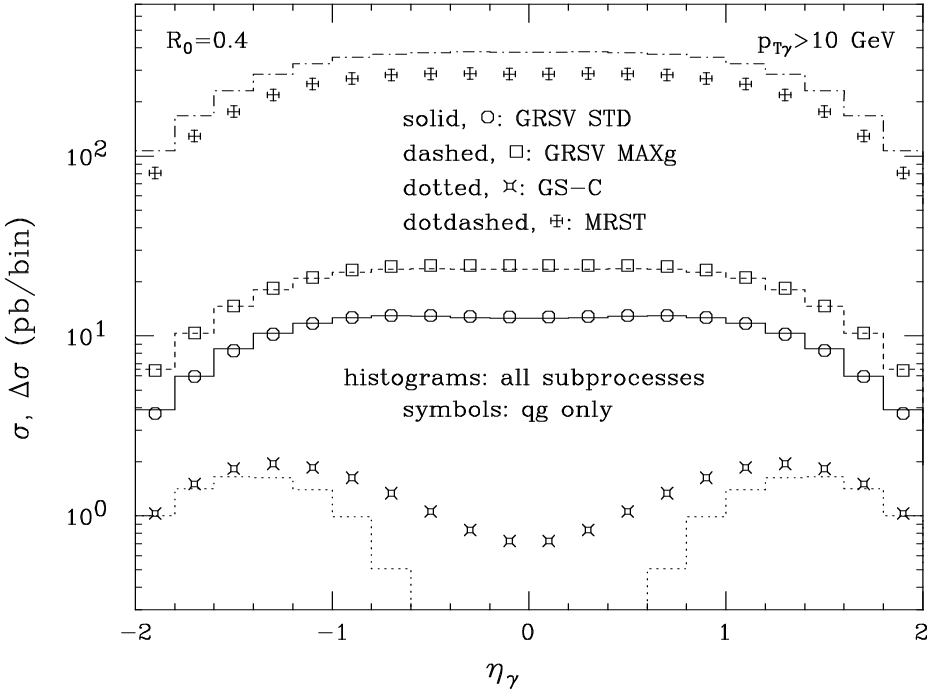,width=0.48\textwidth,clip=} }
\ccaption{}{ \label{fig:qgdom}
$\ptg$ and $\etag$ spectra of isolated photons in polarized and 
unpolarized collisions. The results obtained by retaining only
the contribution of the $qg$-initiated partonic subprocess
are also shown (symbols).
}
\end{figure}                                                              
We now return to the issue of the dominance of the $qg$-initiated
subprocess in prompt-photon production at RHIC. In fig.~\ref{fig:qgdom}
we show the $\ptg$ (left) and $\etag$ 
(right) dependences of the polarized and unpolarized
cross sections, using the three sets of polarized parton densities
employed in the previous figures. As before, the isolation of definition B 
and the default choice of scales have been adopted. 
We have used $\sqrt{S}=200$~GeV
here (we checked that at $\sqrt{S}=500$~GeV we obtain results that
lead us to the very same conclusions); cuts are as before. 
The symbols show in each case the cross section that 
is obtained by keeping only the $qg$-initiated subprocess.
One can clearly see that
in the polarized case, for the two GRSV density sets, the $qg$ subprocess
alone produces a result that is almost identical to the full answer. As we 
have checked, this comes about to some extent because the other subprocesses
all give small contributions, but also because they tend to cancel
one another to a good approximation. This explains why for set
GS-C a different pattern is found: slightly different quark densities
and a vastly different gluon distribution make the cancellation of
the non-$qg$ channels imperfect, and the gluon density is not large
enough to render the signal from $qg$ scattering dominant, except
for large $|\etag|$.  
However, in view of fig.~\ref{fig:asypt}, this finding does not 
really create a problem: if the gluon is indeed as small as embodied
in the GS-C set, the measurement at RHIC will anyway only give
an asymmetry compatible with zero, and we will not be in a
position to actually unfold $\Delta g$ from the data. If,
on the other hand, $\Delta g$ is sizeable, it is an encouraging
result that the polarized cross section provides a very direct measure
of it. Note that in the unpolarized case the $qg$ channel 
is generally responsible for only ${\cal O}(80\%)$ of the cross section.

We finally mention that we have also computed the
asymmetries by isolating the photon according to definition A. In the
case of the GRSV sets, the results are almost identical to those
shown here. In the case of GS-C, some difference can be seen
in the central pseudorapidity region, where the asymmetry tends to
be smaller in the case of definition A. Part of this effect
results from the different scale choices adopted for the two
definitions, as discussed at the beginning of this section.

\section{Isolated-photon-plus-jet observables\label{sec:corr}}

In the production process, the transverse 
momentum of the prompt photon is balanced
by that of the high-$\pt$ outgoing hadrons. It may be decided to neglect the
properties of these hadrons, and to study the inclusive production of
the photon, as we did in section~\ref{sec:si}. On the other hand, the
study of the correlations between the photon and the associated
hadrons gives a more thorough information on the underlying dynamics.
Also, from the experimental point of view, photon-plus-hadron events
can be used as a means of calibrating the hadronic calorimeter. In this
section, we will consider photon-plus-jet observables. Our predictions
are relevant to the STAR experiment~\cite{bland} at RHIC, where one 
of the main goals is indeed to determine $\Delta g$ from 
prompt-photon-plus-jet events. As is customary
in any fixed-order computation in perturbative QCD, our predictions
are given at the parton level (i.e. our jet-finding algorithm deals
with partons and not with hadrons). We will adopt here a
$\kt$-algorithm, namely that proposed in ref.~\cite{ESalg}, with
$D=1$. We will only discuss the case when the photon is isolated
following the prescription B; unless otherwise specified, we will
adopt the isolation parameters given in eq.~(\ref{isolparB}):
$R_0=0.4$, $\epg=1$, $n=1$. Since our computation is based on five-leg
amplitudes, we are able to predict the photon-plus-one-jet observables
at NLO, and the photon-plus-two-jet observables at LO.

\subsection{General features and perturbative stability
\label{subsec:prel}} 
The number of jets that accompany the photon does depend not only
upon the dynamics, but also upon the jet-finding algorithm and 
the kinematical cuts imposed on the jets. This is documented in 
table~\ref{tab:phjetxsec}, where the total rates are presented for 
events satisfying
\beq
\ptg\ge 10~{\rm GeV},\;\;-1\le\etag\le 2,\;\;\;\;\;\;
\ptj\ge 10,12,14~{\rm GeV},\;\;-1\le\etaj\le 2,
\eeq
in the case of polarized and unpolarized $pp$ collisions at $\sqrt{S}=500$~GeV
($\ptj$ and $\etaj$ are the jet transverse momentum and pseudorapidity, 
respectively). The cuts on $\etag,\etaj$ considered here are relevant
to the STAR detector. As one might expect, there is only a small
fraction of events where the photon is accompanied by two jets 
\begin{table}
\begin{center}
\begin{tabular}{|l||c|c|c|c|c|c|c|c|c|} \hline
& \multicolumn{3}{c|}{$\ptj\ge 10$ GeV} 
& \multicolumn{3}{c|}{$\ptj\ge 12$ GeV} 
& \multicolumn{3}{c|}{$\ptj\ge 14$ GeV} 
\\ \hline
& 0-jet & 1-jet & 2-jet 
& 0-jet & 1-jet & 2-jet 
& 0-jet & 1-jet & 2-jet 
\\ \hline\hline
GRSV STD  & 115.1& 168.0& 15.87& 136.7& 151.4& 10.92& 178.0& 113.3& 7.692
\\ \hline
GRSV MAXg & 206.7& 294.7& 29.28& 244.6& 265.8& 20.30& 317.2& 199.1& 14.41
\\ \hline
GS-C      & 27.73& 41.57& 1.121& 37.30& 32.52& 0.604& 49.61& 20.50& 0.312
\\ \hline
MRST$\cdot 10^{-3}$      
          & 11.61& 13.27& 0.730& 14.37& 10.77& 0.471& 18.45& 6.854& 0.310
\\ \hline
\end{tabular} 
\end{center}                                                            
\ccaption{}{\label{tab:phjetxsec}
Total rates (in pb; the entries relevant to unpolarized scattering
have been multiplied by $10^{-3}$) for isolated-photon-plus-jet
events, at $\sqrt{S}=500$~GeV.
}
\end{table}                                                               
(in the case of two-jet events, the transverse-momentum cut is 
applied to both jets; it is clear that the inclusion of radiative
corrections for photon-plus-two-jet observables will not change what
was said before). On the other hand\footnote{By definition, for a given 
row in table~\ref{tab:phjetxsec}, the sum `0-jet'+`1-jet'+`2-jet' 
is the same for each of the three $\ptj$ cuts, and corresponds to
the inclusive isolated-photon rate.}, a sizeable number of
events falls in the class denoted by `0-jets', which is
constituted by those events where the jet(s) do not pass the imposed
transverse-momentum or pseudorapidity cuts; at the lowest transverse-%
momentum cut, this is mainly due to the fact that the pseudorapidity
cut is not symmetric around $\eta=0$.  Since the number of jets is
directly related to the hardness of the event, large differences 
can be seen in the ratio of two-jet over one-jet rates, when different
parton densities are considered. In particular, this ratio is the larger
the slower the gluon density approaches zero for $x\to 1$; while GRSV
STD and GRSV MAXg return almost the same ratio (the shape of their
gluon being basically the same), the result for GS-C is much smaller,
since the gluon density in this case has a dip at intermediate $x$
values. It follows that a first rough piece of information on the
behaviour of the gluon density at large $x$ can be obtained by simply
looking at the {\em total} photon-plus-jet rates.  We also notice that the
result for the ratio of one-jet over two-jet rates
in the case of unpolarized scattering (fourth row
in table~\ref{tab:phjetxsec}) lies in between that of the GRSV sets
and that of the GS-C set, consistently with the fact that the shape of
the MRST gluon density is softer than that of GRSV STD and harder than
that of GS-C. This also implies that the ratio of rates corresponding to
different numbers of jets is not very sensitive to the polarization of
the beams.

\begin{figure}
\centerline{
   \epsfig{figure=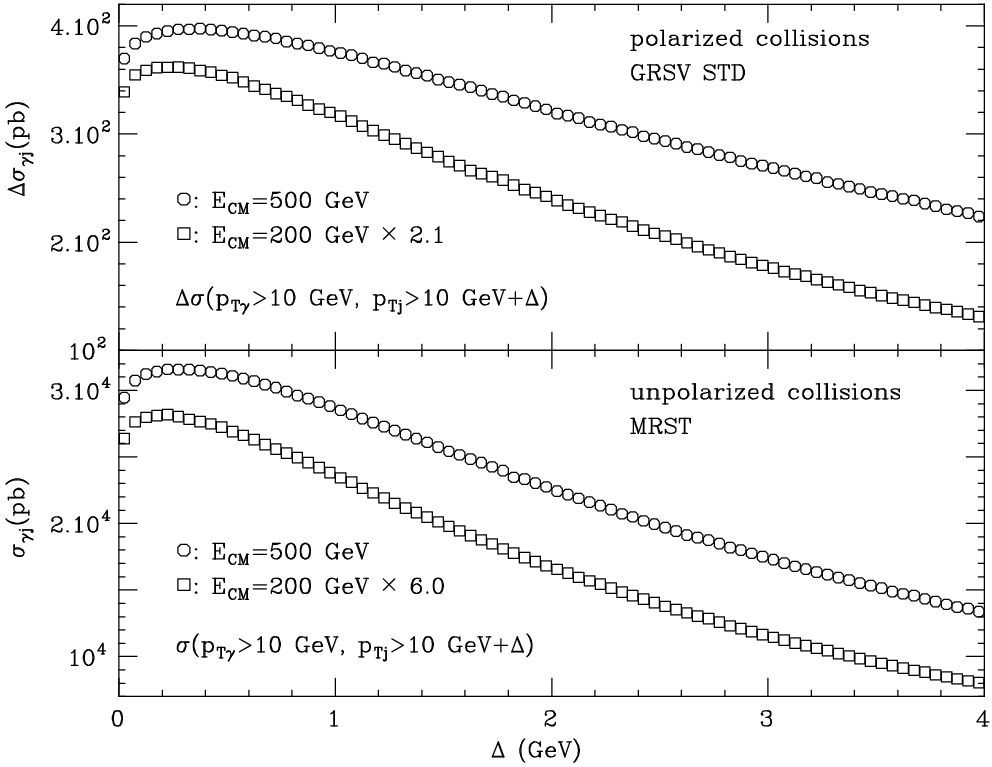,width=0.7\textwidth,clip=}
           }
\ccaption{}{ \label{fig:delta}
Total photon-plus-jet rates, in polarized and unpolarized collisions,
at different centre-of-mass energies.
}
\end{figure}                                                              
We must stress that, in the case of $\ptj\ge~10$~GeV {\it and} 
$\ptg\ge~10$~GeV, the zero-jet and one-jet rates are rather pathological 
in perturbative QCD (on the other hand, their sum is well-behaved). Indeed, 
when equal transverse-momentum cuts are imposed on the photon and the hardest 
jet, large logarithms appear in the cross section, which in principle
should be resummed. Roughly speaking, at any fixed order in perturbation
theory, for `symmetric cuts' the radiation of real gluons cannot 
compensate the large and negative contribution of the virtual diagrams. 
The mechanism is identical to the one that can be observed in two-jet 
correlations, in the case when the two jets have the same minimum 
transverse momentum cut. This matter was discussed at length in 
ref.~\cite{sg2}, to which we refer the reader for more details. 
For illustration, we consider here the total rate (no $\eta$ cuts have 
been applied; these would just change the absolute normalization which is 
of no interest in what follows):
\beq
\sigma_{\gamma j}(\Delta)=
\sigma(\ptg\ge 10~{\rm GeV},\ptj\ge 10~{\rm GeV}+\Delta)
\eeq
as a function of $\Delta$, for both polarized and unpolarized collisions,
at different centre-of-mass energies. 
By definition, the jet is the hardest of the jets of the 
event. The results are displayed in fig.~\ref{fig:delta}. The plots in this 
figure are completely analogous to the ones in fig.~4 of ref.~\cite{sg2}. 
The main point is that a negative slope is here visible at $\Delta=0$, 
implying that the cross section {\em decreases} here with the {\em decreasing}
cut on $\ptg$, clearly signalling a failure of the perturbative expansion. 
We remark, however, that, at variance with the case of jet--jet correlations, 
in the case of isolated-photon-plus-jet production, a value of $\Delta=1$~GeV 
already appears to be perfectly safe. We also remind the reader that, 
even in the case of equal transverse-momentum cuts, the perturbative 
expansion is reliable everywhere except in some corners of the phase space
(examples will be given below). Inspecting fig.~\ref{fig:delta}, we finally 
note that, when going from $\sqrt{S}=200$~GeV to $\sqrt{S}=500$~GeV, 
the cross section increases much more in the unpolarized than in the 
polarized case. This implies that, as in the case of inclusive observables
discussed in the previous section, at fixed final-state kinematical 
variables the asymmetries for photon-plus-jet 
observables are larger at the smaller centre-of-mass energies.

\begin{figure}
\centerline{
   \epsfig{figure=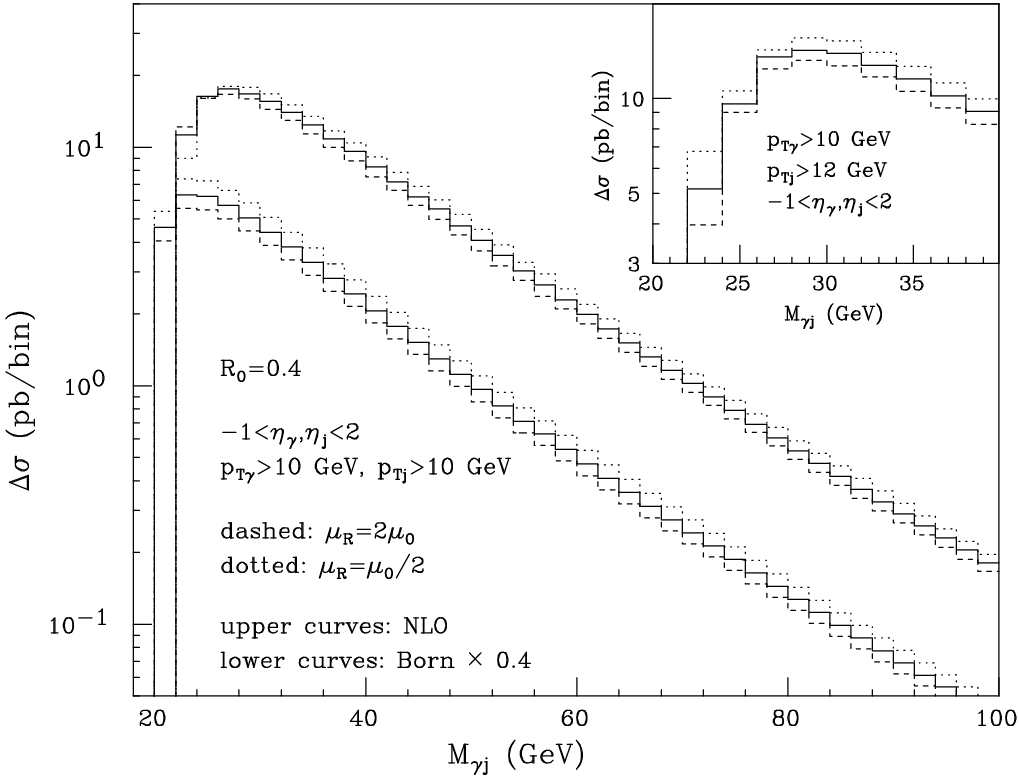,width=0.7\textwidth,clip=}
           }
\ccaption{}{ \label{fig:massgj}
Renormalization-scale dependence of the invariant mass spectrum of the 
photon--jet system, at the Born and NLO levels. The result for unequal 
lower transverse-momentum cuts is also shown (inset).
}
\end{figure}

We now turn to the issue of the perturbative stability of our
results for correlations between photon and jets. Since we are
interested in NLO predictions, we will only consider photon-plus-%
one-jet quantities; in the case when two jets are present in
the event, only the hardest jet is retained. As in 
subsection~\ref{subsec:si1}, we will assume that we obtain 
a (relatively sound) indication of the stability of the cross 
sections if the variations induced by changing the 
renormalization and factorization scales with respect to their 
default values are small. In fig.~\ref{fig:massgj} we present the 
result for the invariant mass distribution of the photon-jet
system, in the case of polarized collisions at $\sqrt{S}=500$~GeV.
The pseudorapidities of both the photon and the jet are required to be
in the range \mbox{$-1\le\eta\le 2$}, and we impose $\ptg\ge 10$~GeV
and $\ptj\ge 10$~GeV. The results of both the NLO computation (upper
curves) and Born computation (lower curves, which have been rescaled in
order to make them clearly distinguishable from the NLO ones) are displayed.
Similarly to the case of the inclusive transverse momentum distribution
of isolated photons (cf. fig.~\ref{fig:ptspectrumB}), in most of the 
range in $M_{\gamma j}$ the change of cross section induced by a 
variation of the renormalization scale is of the order of 10\% at 
NLO, and larger at the Born level. However, a quite dramatic
effect is seen at threshold, when radiative corrections
are included: the cross section in the first bin becomes negative, 
and the scale dependence displays a pathological behaviour in this 
range, the cross section becoming smaller for decreasing renormalization 
scales. This effect is exactly a consequence of the fact that the minimum 
transverse-momentum cuts on the photon and the jet are equal.
Indeed, in the inset of fig.~\ref{fig:massgj} we show (in the
threshold region) the invariant mass of the photon--jet system 
in the case when $\ptj\ge 12$~GeV. It is obvious that here the
scale dependence is as expected, and the cross section in the first 
bin (although not visible in the figure) remains positive. 
However, the first two bins show a scale sensitivity 
comparable to the one of the Born result. In fact, close to the 
threshold the NLO result is effectively a LO one, since the threshold 
at the Born level is in this case 
at \mbox{$M_{\gamma j}=24$~GeV}. In the invariant
mass range not close to threshold, the scale dependence
in the case of unequal transverse-momentum cuts is practically
identical to the one displayed in the main body of fig.~\ref{fig:massgj}.
We also studied the factorization scale dependence of the 
invariant mass distribution. In the region not close to the 
threshold, there is a clear improvement when going from LO
to NLO; again, the results for equal and unequal transverse-%
momentum cuts are very similar. At threshold, the same 
considerations as given above apply. As in the case of the 
single-inclusive photon transverse-momentum spectrum displayed in
fig.~\ref{fig:ptspectrumB}, the cross section for larger (smaller)
factorization scales is larger (smaller) than the default
one at small invariant masses, while it is smaller (larger)
than the default for large invariant masses. This behaviour
is almost entirely due to scale dependence of the parton 
densities, as already discussed in the single-inclusive case. 

We performed a thorough study of the renormalization and 
factorization scale dependence of many photon-jet observables.
In particular, we considered the photon (jet) transverse momentum and 
pseudorapidity distributions, when cuts on the recoiling jet (photon)
are imposed, as suggested in ref.~\cite{lg1}. Among the photon-jet
correlations, we considered the transverse momentum of the pair
$p_{\sss T}^{(\gamma j)}$, the azimuthal distance in the transverse 
plane $\Delta\phi_{\gamma j}$, the distance in the $\eta$--$\phi$ plane
$\Delta R_{\gamma j}$, and the variables
\beq
x_1=\frac{\ptg e^{\etag}+\ptj e^{\etaj}}{\sqrt{S}},\;\;\;\;\;\;\;\;
x_2=\frac{\ptg e^{-\etag}+\ptj e^{-\etaj}}{\sqrt{S}},
\eeq
which, at the Born level, coincide with the Bj\o rken-$x$ values of the 
incoming partons. In all these cases, a reduction in the relative size 
of the scale dependence is seen at the NLO with respect to the Born 
level, whenever such a comparison is meaningful (that is whenever 
the Born contribution is already present at the level of a $2\to 2$
scattering process): a change of the scales within the limits as 
above induces a variation of the results of about 10\% or less. In 
the regions of the phase space where the partonic contributions
start at the $2\to 3$ level (for example, $p_{\sss T}^{(\gamma j)}>0$,
$\Delta\phi_{\gamma j}<\pi$, $\Delta R_{\gamma j}<\pi$), our NLO
results have a scale dependence larger than elsewhere, of the order of
15\% to 20\%, since they are effectively LO. Finally, as in the 
case of the invariant-mass distribution, there are corners of the phase 
space where the perturbative results are not reliable in the case of 
equal transverse-momentum cuts. Among those, the case of $\Delta 
R_{\gamma j}=\pi$ is particularly interesting, since here the Born 
threshold falls inside the range available at NLO. This case has been
described, on general grounds, in ref.~\cite{CW}.

\subsection{Spin asymmetries\label{subsec:pjphen}}
\begin{figure}
\centerline{
   \epsfig{figure=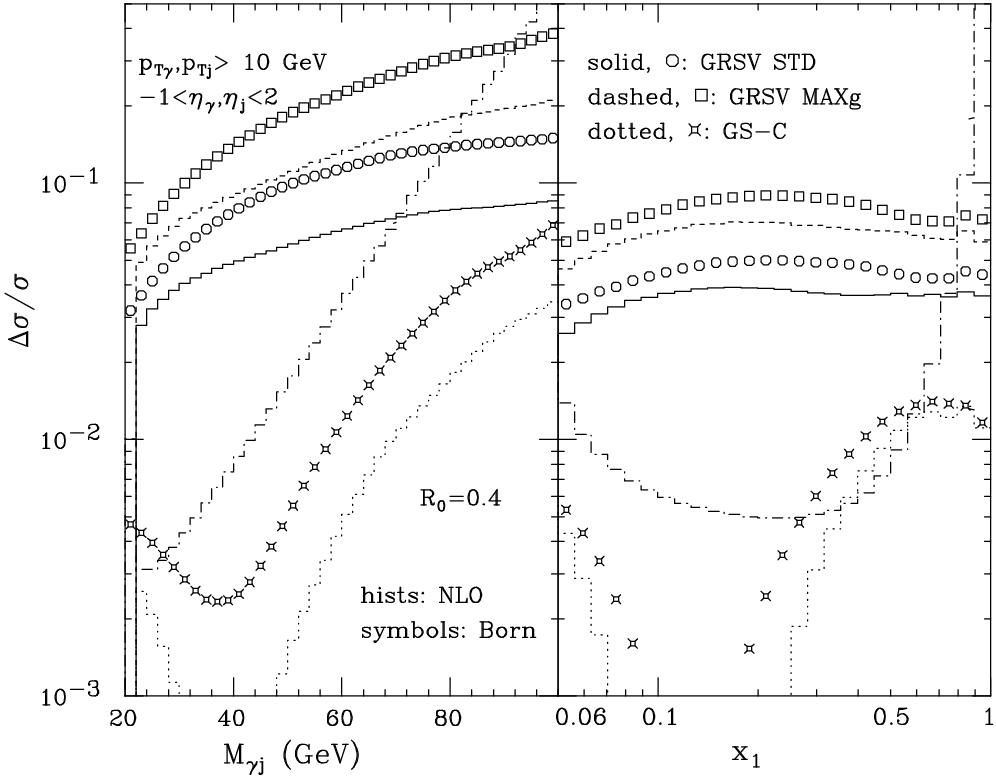,width=0.7\textwidth,clip=}
           }
\ccaption{}{ \label{fig:asycorr}
Asymmetries at $\sqrt{S}=200~GeV$, as functions of invariant mass 
(left) and $x_1$ (right). The NLO (histograms) and Born (symbols) 
results are both shown. The minimum observable asymmetry is displayed 
by the dot-dashed histogram.
}
\end{figure}

We now turn to the study of asymmetries for photon-plus-jet
cross sections. We follow the procedure of subsection~\ref{subsec:si2}, 
namely we study the dependence of the asymmetries on the choice of
polarized parton densities, at the Born and NLO levels.
Here, we restrict ourselves to $\sqrt{S}=200$~GeV.
We verified that the pattern when going to $\sqrt{S}=500$~GeV
is similar to the one already described in the preceding section;
namely, at fixed final-state kinematics
we get smaller asymmetries, with however also smaller minimally 
observable asymmetries. In fig.~\ref{fig:asycorr} we show the asymmetries 
as functions of the invariant mass (left) and $x_1$ (right). 
The photon-plus-jet events have been selected by imposing equal transverse-%
momentum cuts on the photon and on the hardest jet ($\pt>10$~GeV; 
pseudorapidities are restricted to the range \mbox{$-1<\eta<2$}). 
As discussed previously, this choice only affects the threshold region 
of the invariant mass, where our predictions should not be considered as
reliable. As in the case of single-inclusive quantities, the results 
for GRSV STD and GRSV MAXg are pretty similar in shape, although sizeably 
different in normalization. On the other hand, GS-C has a clearly 
distinguishable signature, showing a dip at intermediate values of 
the invariant mass and in the region around $x_1=0.1$.
It is very easy to trace the origin of this behaviour back to
the shape of the GS-C gluon. The NLO results are smaller than
those at LO, as they already were in the case of inclusive observables.
The difference between Born and NLO results is not big at small
invariant masses and in the whole $x_1$ range, while it grows larger
in the tail of the invariant-mass distribution, since the $K$-factor 
of the unpolarized cross section is larger in this region than that 
of the polarized cross section.
Figure.~\ref{fig:asycorr} also presents the minimally observable asymmetry 
(dot-dashed histograms), calculated according to eq.~(\ref{minasy}). 
A bin size of 2~GeV has been chosen for $M_{\gamma j}$, and of $0.05$
for $\log x_1$; as before, ${\cal L}=100$~pb$^{-1}$ and
$\ep=P=1$. It is apparent that, if the polarized densities are
as predicted by the GS-C set, the measurement of the asymmetry 
at RHIC will produce a result compatible with zero, even
if quite large integrated luminosities are attained. This result
is consistent with what we already found in section~\ref{subsec:si2};
however, in the case of single-inclusive variables the situation
appeared to be slightly better, when the enlarged pseudorapidity
range \mbox{$-1<\eta<2$} was considered.
 
\begin{figure}
\centerline{
   \epsfig{figure=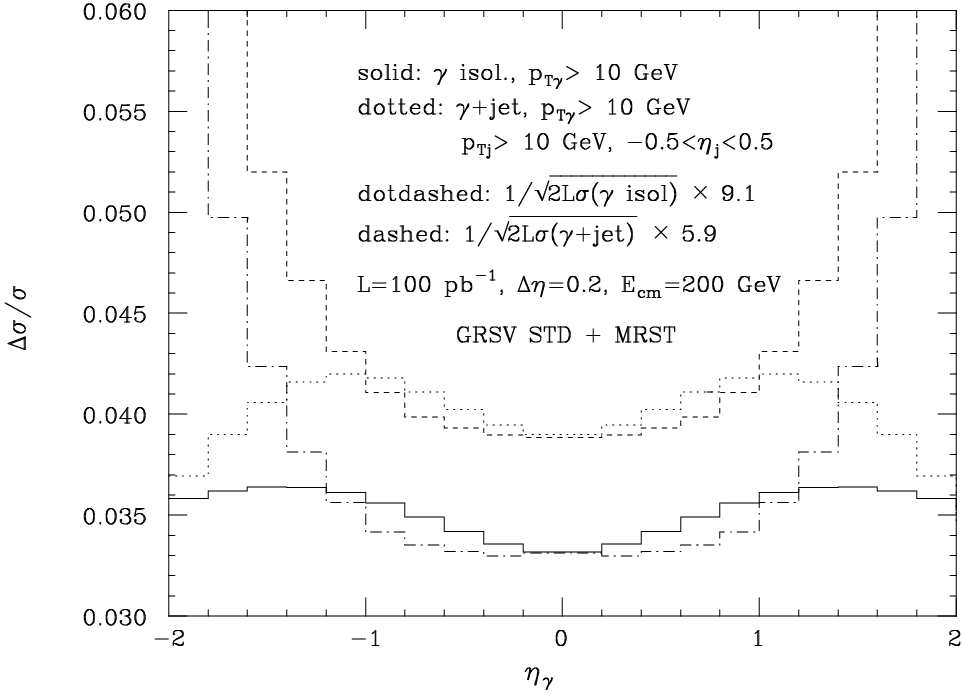,width=0.7\textwidth,clip=}
           }
\ccaption{}{ \label{fig:etacomp}
Asymmetries at $\sqrt{S}=200~GeV$, as functions of photon pseudorapidity, 
with (dotted) and without (solid) a transverse-momentum cut on the 
recoiling jet. The corresponding minimally observable asymmetries 
(dashed and dot-dashed, respectively) have been rescaled in order to 
coincide with the respective asymmetries at $\etag=0$.
}
\end{figure}

In this context, we would like to comment on the findings of 
ref.~\cite{lg1}, where it was observed that, by looking at
photon-plus-jet events, instead of considering only the inclusive
variables of the photon, one gets larger asymmetries. Also,
photon-plus-jet observables enhance the sensitivity to the shape of
the parton densities and can be used for a more straightforward
deconvolution of $\Delta g$ from data~\cite{bland}. Although we agree with 
these observations, we doubt that photon-plus-jet correlations will
give us a better chance of measuring the gluon density than inclusive 
observables. In fact, there is in practice the problem that the (theoretical)
minimally observable asymmetry is larger in the case of photon-plus-jet
quantities than in the case of inclusive-photon quantities. The
situation is summarized in fig.~\ref{fig:etacomp}, for the case of
the $\etag$ spectrum with $\ptg\ge 10$~GeV. In the case of the 
photon-plus-jet observable, the following cuts have been imposed on the 
jet: $\ptj\ge 10$~GeV, $\abs{\etaj}\le 0.5$. From the figure, it is 
apparent that, in spite of the fact that the asymmetry is increased 
when cutting on the jet variables, the measurement would be more difficult, 
since the minimally observable asymmetry is enhanced by a larger 
factor with respect to the asymmetry. Of course, this is a purely 
theoretical estimate, which assumes that the experimental efficiency 
is the same in the case of inclusive-photon or photon-plus-jet events 
(here taken to be 
equal to 1). This is probably unrealistic, but it seems unlikely that the
photon-plus-jet efficiency will be higher than that for photon-tagging
only. Therefore, without any detailed study at the detector level,
it seems improbable that the photon-plus-jet asymmetries will be the
preferred tool for pinning down the polarized gluon density. 
We finally also have to add that in ref.~\cite{lg1} the photon
transverse momentum was constrained in a bin around $\ptj=10$~GeV 
of width $1$ GeV, instead of having $\ptg>10$~GeV as in 
fig.~\ref{fig:etacomp}. In view of the discussion relative to 
fig.~\ref{fig:delta}, the kinematical constraints imposed in
ref.~\cite{lg1} appear to be more problematic from the perturbative
point of view than those adopted in this paper for producing
fig.~\ref{fig:etacomp}.

\section{Conclusions\label{sec:concl}}

We have performed a phenomenological study of prompt-photon production 
by polarized colliding proton beams at RHIC. Our main motivation has
been to put in practice a recently proposed alternative way of isolating the
photon from the hadronic background, which eliminates the ill-understood
fragmentation contribution from the cross section, so that only the
direct component is left. In this way, an optimally clean photon
signal results. This has also enabled us to perform the first 
fully consistent next-to-leading order calculation for polarized 
prompt-photon production. 

We have compared the cross sections and asymmetries for the new isolation
with those obtained for a traditional cone-type isolation as used 
hitherto in unpolarized collider prompt-photon experiments. We find only small 
differences, implying that using the new `clean' isolation is not accompanied 
by a reduction in the number of events. 

Performing a consistent next-to-leading order study, we were in a
position to examine the perturbative stability of the cross sections by
studying their dependence on the factorization and renormalization
scales. We find that the scale dependence is significantly reduced
when going from the Born level to the next-to-leading order, and that
the scale dependence is moderate at next-to-leading order. This generally 
holds true for both inclusive isolated-photon production and for 
isolated-photon-plus-jet production. 

We have presented phenomenological results for the spin asymmetries
and their expected statistical errors in the RHIC experiments PHENIX
and STAR. Consistently with previous studies, the results are highly 
sensitive to the size of the 
spin-dependent gluon density $\Delta g(x,Q^2)$, and it turns out that,
unless $\Delta g$ is very small, it should clearly be possible
to rather accurately determine it in the region $0.04<x<0.25$, at scales
of the order of the transverse momentum of the photon, 
10~GeV~$<\ptg<$~30~GeV. In this context, we find that as far as 
statistics is concerned, inclusive-photon measurements seem to be somewhat 
more favourable than photon-plus-jet ones. 

\section*{Acknowledgements}
We are happy to thank G. Ridolfi for his collaboration at an early
stage of this work. We are also grateful to N. Saito for useful
information.
The work of S.F. is supported in part by the EU Fourth 
Framework Programme `Training and Mobility of Researchers', Network
`Quantum Chromodynamics and the Deep Structure of Elementary Particles',
 contract FMRX-CT98-0194 (DG 12 - MIHT).

\end{document}